\documentclass[final,5p,times,twocolumn]{elsarticle}

\usepackage{latexsym}
\usepackage{amssymb, amsmath, bm, physics}
\usepackage{subcaption}
\usepackage{mathptmx}
\usepackage{flushend}
\usepackage[colorlinks,citecolor=blue,urlcolor=blue,linkcolor=blue]{hyperref}
\usepackage{aas_macros}
\usepackage{orcidlink}
\DeclareMathOperator{\Si}{Si}
\usepackage{xcolor}

\biboptions{sort&compress}
\journal{Physics of the Dark Uiverse}

\begin{document}

\begin{frontmatter}

\title{Modified cosmology from quantum deformed
entropy}

\author[1]{S. Jalalzadeh %\corref{cor1} 
\orcidlink{0000-0003-4854-2960}}
\ead{shahram.jalalzadeh@ufpe.br}
\address[1]{Departamento de F\'{i}sica, Universidade Federal de Pernambuco, Recife-PE, 50670-901, Brazil}
\cortext[cor1]{Corresponding author}
\author[2]{H. Moradpour
\orcidlink{0000-0003-0941-8422}}
\ead{h.moradpour@riaam.ac.ir}
\address[2]{Research Institute for Astronomy and Astrophysics of Maragha (RIAAM), P.O. Box 55134-441, Maragheh, Iran}
\author[3]{P.V. Moniz \orcidlink{0000-0001-7170-8952}}
\ead{pmoniz@ubi.pt}
\address[3]{Departamento
de F\'isica and Centro de Matem\'atica e Aplica\c c\~oes (CMA - UBI),  Universidade da Beira Interior,  
%Rua Marqu\^es d’\'Avila e Bolama,  
6200 Covilh\~a,  Portugal}

\begin{abstract}
In Ref. \cite{Jalalzadeh:2022rxx}, Jalalzadeh established that the thermodynamical entropy of a quantum-deformed black hole with horizon area $A$ can be written as  $S_q=\pi\sin\left(\frac{A}{8G\mathcal N} \right)/\sin\left(\frac{\pi}{2\mathcal N} \right)$, where $\mathcal N=L_q^2/L_\text{P}^2$, $L_\text{P}$ being the Planck length and $L_q$ denoting, generically, the q-deformed cosmic event horizon distance $L_q$. Motivated by this, we now extend the framework constructed in \cite{Jalalzadeh:2022rxx} towards the Friedmann and Raychaudhuri equations describing spatially homogeneous and isotropic universe dynamics. Our procedure in this paper involves a twofold assumption. On the one hand, we take the entropy associated with the apparent horizon of the Robertson--Walker universe in the form of the aforementioned expression. On the other hand, we assume that the unified first law of thermodynamics, $dE=TdS+WdV$, holds on the apparent horizon. Subsequently, we find a novel modified cosmological scenario characterized by quantum-deformed (q-deformed)    Friedmann and Raychaudhuri equations containing additional components that generate an effective dark energy sector. Our results indicate an effective dark energy component, which can explain the Universe’s late-time acceleration. Moreover, the Universe follows the standard thermal history, with a transition redshift from deceleration to acceleration at $z_\text{tran}=0.5$. More precisely, according to our model,  at a redshift of $z = 0.377$,  the effective dark energy dominates with a de Sitter universe in the long run. We include the evolution of luminosity distance, $\mu$, the Hubble parameter, $H(z)$, and the deceleration parameter, $q(z)$, versus redshift. Finally, we have conducted a comparative analysis of our proposed model with others involving non-extensive entropies.
\end{abstract}

%\begin{keyword}
% Holographic dark energy\sep Modified cosmology \sep q-deformation
%\end{keyword}

\end{frontmatter}

%\linenumbers


\section{Introduction}\label{Int}
Entropy is a fundamental quantity, constituting a significant 
concept.  It is desirable that a generalized definition of entropy may apply to all physical systems; otherwise,  it may mean that we do not grasp what physical entropy is. In fact,  as quantum mechanics, quantum field theory, and quantum gravity have developed,  entropy has not been cast as a universal property or a generic form, but instead, it has been presented
within many contexts, depending on the particular physical system under investigation. For instance, one of the key findings in theoretical physics is that a black hole (BH) is associated with blackbody radiation, which has a distinct temperature and Bekenstein--Hawking entropy \cite{Bekenstein:1973ur,Hawking:1975vcx}. Unlike classical thermodynamics, where the entropy is directly related to the system's volume and represents an extensive quantity, the Bekenstein--Hawking entropy is proportional to the horizon's area.

Interestingly enough, it has also been shown that the field equations of general relativity can be obtained as a thermodynamic equation of state \cite{Jacobson:1995ab}, a result also confirmed in modified gravity \cite{Eling:2006aw}. Parallel to this achievement signalling a deep connection between gravity and thermodynamics and hence statistical mechanics, it has also been addressed that quantum aspects of gravity may trigger gravity to show intrinsic non-extensive features 
%from itself 
\cite{Moradpour:2019yiq, Shababi:2020evc}. Indeed, the long-range nature of gravity is another similar reason to study cosmology in the framework of {\textit{generalized statistics}} and corresponding thermodynamics \cite{moradpour:ijtp,Tavayef:2018xwx, SayahianJahromi:2018irq, Moradpour:2018ivi, Moradpour:2017ycq, Moradpour:2020dfm}. {Employing those statistics as mentioned earlier} may also provide an origin for dark matter and MOND theory \cite{Moradpour:2017fmq, Senay:2020zya}. The applications are not limited to these areas, and it seems there is hope to solve the problems of BH physics, astrophysics, and high-energy physics using the generalized statistics framework. \cite{Tsallis:2014hta, Esquivel:2009te, PAVLOS2015113, Moradpour:2022Ent, Ourabah:2023byw, Luciano:2022Ent, Moradpour:2021soz}.

Recent literature has proposed several 
definitions of entropy based on non-additive statistics, {such as the Tsallis \cite{Tsallis:1987eu,2013EPJCT},} Rényi \cite{renyi1961}, and Barrow \cite{Barrow:2020tzx} entropies, in addition to the Bekenstein--Hawking one, due to the elusive nature of entropy. Moreover, the authors of Refs. \cite{Jalalzadeh:2021gtq,Jalalzadeh:2022uhl} proposed a fractional-fractal entropy that could encode the random fractal features of a BH horizon surface that resulted from fractional quantum gravity effects. The Sharma--Mittal entropy \cite{SayahianJahromi:2018irq}, the Kaniadakis entropy \cite{Hernandez-Almada:2021rjs,Lymperis:2021qty,Luciano:2022knb}, the entropy in the setting of Loop Quantum Gravity \cite{Majhi:2017zao,Liu:2021dvj}, the quantum cosmology \cite{Rashki:2014noa,Jalalzadeh:2014jea,Jalalzadeh:2017jdo,Bina:2010ir} and the entropy in fractional quantum gravity-cosmology \cite{Jalalzadeh:2022uhl,Jalalzadeh:2021gtq,Jalalzadeh:2020bqu} 
constitute some more well-known entropies.

Furthermore, a novel entropy formula for a BH was recently proposed by one of us \cite{Jalalzadeh:2022rxx}, based on the quantum deformation (or q-deformation) approach to quantum gravity. Very briefly, let us {explain} this entropy.

 According to various quantum gravity proposals, the BH horizon area, $A_n$, may be quantized, and the appropriate eigenvalues are given by \cite{Bekenstein:1974jk}
\begin{equation}
    \label{H1}
    A_n=\gamma{\,} L_\text{P}^2{\,}n,~~~~n=1,2,3,...~,
\end{equation}
where $\gamma$ is a model-dependent dimensionless constant of order one, and $L_\text{P}=\sqrt{G}$ is the Planck length. 
{Since Bekenstein's groundbreaking work, the literature has been enriched with a multitude of contributions that reinforce the conjecture of the area spectrum (\ref{H1}). These contributions are diverse and multifaceted, encompassing various considerations such as information theory (as outlined in \cite{Danielsson:1993um,Bekenstein:1995ju}), arguments from string theory \cite{Mazur:1987jf}, or the periodicity of time \cite{Mazur:1986gb,Mazur:1987sg,Bina:2010ir}. Furthermore,  contributions extended from  the quantization of a dust collapse using Hamiltonian methods \cite{Peleg:1995gg,Nambu:1987dh}, can also be mentioned.}
However, a BH can be considered physically embedded in the 
{Universe, characterized by event horizon $A_\text{U}$}. 
As a result, the Schwarzschild event horizon of a BH cannot be larger than the cosmic event horizon of the Universe, $A_\text{U}$. The usual quantization methods with a spectrum given by Eq. (\ref{H1}) must consistently meet
this reasoning. It has to be adjusted to be bounded from above. Quantum deformation of a model (using quantum groups) is one 
such method, allowing retrieval of the dimension of a Hilbert space into a finite value, assuming that the deformation parameter is a root of unity \cite{Frohlich:1993km,Alvarez-Gaume:1988izd,Pasquier:1989kd,Grosse:2000gd,Steinacker:2000yk,Livine:2016vhl,Pouliot:2003vt}. 

In this work, we propose to extend the quantum-deformed (q-deformed) entropy idea, applying the first law of thermodynamics to the Universe's {horizon,  adapting to a cosmological scenario} the thermodynamical entropy of a quantum-deformed  BH with horizon area $A$ presented in \cite{Jalalzadeh:2022rxx}. 
This allows us to construct a q-deformed modified version of the Friedmann and Raychaudhuri equations. In particular, there will be newly added components whose dynamical implications motivate our investigation. {To make clear our reasoning, let us add the following. The quantum-deformed black hole is a model constructed from the quantum Heisenberg--Weyl group. Quantum groups provide us with more complex symmetries than the classical Lie algebras, which are included in the former as a specific case. This indicates that quantum groups may be appropriate for describing symmetries of physical systems that lie beyond the scope of Lie algebras. Moreover, q-deformed models have a significant advantage because their corresponding Hilbert space is finite-dimensional when $q$ is a root of unity \cite{Klimyk:1997eb}. This implies that using quantum groups with a deformation parameter of the root of unity is useful for constructing models with a finite number of states. These models can be used to explore applications in quantum gravity and quantum cosmology that adhere to the holographic principle and UV/IR mixing to solve the CC problem \cite{Jalalzadeh:2017jdo}.}

{Last but not least, within modified gravity settings, it may be deemed necessary to elucidate the principal rationale behind the development of alternative models for the standard model of cosmology despite its achievement and conformity with observational data. The standard model of cosmology, though successful in its own right, is plagued by a fundamental and unresolved difficulty known as the cosmological constant (CC) problem. This enigma has long baffled cosmologists and remains a significant obstacle in advancing the field.}

{Numerous theoretical physicists expressed their reluctance to acknowledge the CC as a feasible justification for the accelerated expansion of the universe due to the fact that the anticipated value of CC from particle physics is $\rho_\Lambda\simeq M_\text{P}^4\simeq (10^{18}~\text{GeV})^4$, a value that differs significantly from the astronomical limit for CC, which is $\rho_\Lambda\simeq (10^{-3}~\text{eV})^4$ --roughly $10^{123}$ times less than expected. The CC is regarded as the zero-point energy with a UV cutoff scale, such as the Planck scale or the supersymmetry breaking scale, from an effective field theory (EFT) perspective, whereas from a cosmological perspective, it is an IR scale problem that affects the entire universe's large-scale structure. As a result, the CC issue appears to contravene our preconceived notion of separating UV and IR scales, which is the foundation of EFT. The CC can be interpreted as both the zero point energy and the scale of the observed Universe, which contradicts the concept of local quantum fields. This suggests a mixing between the local UV and global IR physics. Some physicists argue that the CC problem is essentially a quantum gravity and quantum cosmology problem \cite{12501,Sheikh-Jabbari:2006czd,Jalalzadeh:2022dlj,Jalalzadeh:2017jdo}. Therefore, a candidate theory for quantum gravity must provide a classical continuum spacetime geometry at macroscopic scales with a global IR cutoff (CC) while also incorporating quantum corrections at the local UV scale. This approach would allow for a better understanding of the complex nature of the CC problem. It is commonly assumed that there exists a solution to the problematic ``old'' CC problem. This solution would result in the vacuum energy being exactly zero and radiatively stable.}

{With the current conditions, it is evident that any suggested substitute for the standard model of cosmology must elude the CC problem. For example, consider the replacement model with several free parameters. By adjusting the free parameters of the model, we may achieve an even more precise fit than the standard model of cosmology. However, the issue arises when we attempt to explain these free parameters' origin and physics in our model, which will involve a comparable problem. In section \ref{sec5} of the article, we will provide an example to revisit this case.}

Our paper structure is as follows.
In the next section, we summarize with minimal detail q-deformed BH entropy. Section \ref{FR} examines the applicability of the previously mentioned procedure
in cosmology. Consequently, this enables us to advance a new modified scenario built
from the q-deformed entropy.
Section \ref{Cosmos} examines the cosmological consequences of the additional components in the q-deformed Friedmann and Raychaudhuri equations, concentrating on the dark energy density and equation-of-state (EoS) parameters. {In section \ref{sec5}, a comparative analysis has been conducted between our proposed model and others previously established in the field of entropic cosmology, with a particular emphasis on examining the similarities and differences between ours and the  Barrow model in 
\cite{Saridakis:2020zol}}. Finally, in Section \ref{Con}, we will examine our findings.

\section{Entropy of a q-deformed Schwarzschild BH}\label{sec1}
Let us briefly go through a few significant results about the q-deformation quantization and entropy of the Schwarzschild BH with mass $M$. Following Louko in Ref. \cite{Louko}, we start with the reduced action of a Schwarzschild BH given by
\begin{equation}\label{1-6}
S=\int \Big\{P_M\dot M-H(M)\Big\}dt,
\end{equation}
where $H(M):=M$ is the reduced Hamiltonian, and  $P_M$ is the canonical conjugate momenta of the BH mass $M$.
The solutions of the field equations are $M=\text{const}.$ and $P_M=-t$. 
The constancy of mass $M$ follows Birkhoff's theorem, which states that the mass is the only time-independent and coordinate-invariant solution. Furthermore, $P_M$ represents the asymptotic time coordinate at the spacelike slice \cite{Kuchar1}. {Thus, $M$ contains all the pertinent information regarding the local geometry of the classical solutions. On the other hand, the conjugate momenta, $P_M$, is equivalent to the disparity of the asymptotic Killing times between the left and right infinities on a constant $t$ hypersurface \cite{Louko}, following the convention where the Killing time at the right (left) infinity increases towards the future (past). As a result, it does not contain any information regarding the local geometry. However, it instead contains information regarding securing the spacelike hypersurfaces at the two infinities \cite{Louko}.}

{Moreover, it is customary to restrict the mass manually and the corresponding momenta to the range $|P_M|<\pi M/M_\text{P}^2$, where $M_\text{P}=1/\sqrt{G}$ is the Planck mass. This implies that the Minkowski time on the asymptotic right-hand side of each classical solution only lies within an interval of length $2\pi M/M_\text{P}^2$, centered around a value that is diagonally opposite to the non-evolving left end of the hypersurfaces in the Kruskal diagram. With respect to the time parameter $t$, each classical solution is exclusively defined for an interval of $-\pi M/M_\text{P}^2<t-t_0< \pi M/M_\text{P}^2$, where $t=t_0$ represents the hypersurface whose two asymptotic ends are diagonally opposite.
As previously stated, in light of Euclidean quantum gravity \cite{1979grec.conf..746H} (where the situation is similar to finite temperature quantum
field theory, when the time is Euclideanized \cite{Blasone:2018ynq}), we posit that the conjugate momentum, $P_M$, serves as a temporal measure and, therefore, must exhibit periodicity with a period inverse to the Hawking temperature $T_\text{H}=M^2_\text{P}/8\pi M$ \cite{Das,Li}. This guarantees the absence of a conical singularity in the two-dimensional Euclidean section close to the black hole horizon. However, it is important to note that the aforementioned identification results in a physical phase space that constitutes a wedge extracted from the complete $(M, P_M)$ plane. The $M$ axis and the $P_M=1/{T_\text{H}}$ line bound this wedge. Therefore,}
%Since $P_M$ plays the
%r%ole of time, it should be periodic in which the period is the inverse Hawking temperature $T_\text{H}=M^2_\text{P}/8\pi M$  \cite{Das,Li}, 
\begin{equation}\label{1-7}
P_M\sim P_M+\frac{1}{T_\text{H}}.
\end{equation}
The above boundary condition verifies that there is no conical singularity in the $2D$ Euclidean section.

Also, it indicates that the phase space is a wedge cut out from the full  $(M, P_M)$ phase space, bounded by the {mass} axis and the line $P_M=1/T_\text{H}$ \cite{Med}. Hence, according to the references \cite{Louko,Bar1}, one could 
make the following canonical transformation $(M, P_M ) \rightarrow (x, p)$, which  simultaneously opens up the phase space and also incorporates the periodicity condition (\ref{1-7})
\begin{equation}
%\begin{split}
x=\sqrt{\frac{A}{4\pi G}}\cos(2\pi P_MT_\text{H}),\,\,\,\,\,\,
p=\sqrt{\frac{A}{4\pi G}}\sin(2\pi P_MT_\text{H}),
%\end{split}
\end{equation}
where $A=16\pi M^2/M_\text{P}^4$ is the BH horizon area. From the above canonical transformations, one immediately finds the horizon area in terms of $(x,p)$
\begin{equation}\label{1-8}
A=4\pi L_\text{P}^2\Big(x^2+p^2\Big),
\end{equation}
where $L_\text{P}=1/M_\text{P}$ is the Planck length.

Let us define the ladder operators, $\{a_-,a_+\}$,  by
\begin{equation}\label{new2}
a_\pm=\frac{1}{\sqrt{2}}\Big(\pm\frac{d}{dx}+x\Big).
\end{equation}
The pairs of operators $a_\pm$ act on states as the following form
\begin{equation}\label{new3}
a_+|n\rangle=\sqrt{n+1}|n+1\rangle,~~~~a_-|n\rangle
=\sqrt{n}|n-1\rangle.
\end{equation}
This gives us the possibility to rewrite the area operator (\ref{1-8}) of the BH in terms of ladder operators
\begin{equation}\label{3-12nona}
A=4\pi L_\text{P}^2\Big(a_+a_-+a_- a_+\Big).
\end{equation}
%This equation shows that the event horizon's area can be written as the Hamiltonian of a simple harmonic oscillator with the mass $m$ and the angular frequency $\omega$ given by $m=1/\omega$. 
Therefore,  the area of the event horizon and the mass spectrum \cite{Louko} are
\begin{equation}\label{1-10}
%\begin{array}{cc}
A_n={8\pi L_\text{P}^2}\Big(n+\frac{1}{2}\Big),~~~~
M_n=\frac{M_\text{P}}{\sqrt{2}}\sqrt{n+\frac{1}{2}},
%\end{array}
\end{equation}
where $n$ is an integer.
Expressions (\ref{1-10}) give the well-known result: Hawking radiation takes place when the BH jumps from a higher state $n+1$ to a lower state $n$, in which the difference in quanta is radiated away. Also, they show that the BH does not evaporate completely, but a Planck-size remnant is left over at the end of the evaporation process.

In 1974,  Hawking \cite{Hawking} showed that due to quantum fluctuations, BHs emit blackbody radiation, and the corresponding entropy is one-fourth of the event horizon area, namely $A=16\pi G^2M^2$.
Following  Refs. \cite{Area1a} and \cite{Xiang}, let us assume that Hawking radiation of a massive BH, i.e., $M \gg M_\text{P}$ and  $n\gg 1$, is 
emitted when the BH system spontaneously jumps from the state $n+1$ into the closest state level, i.e.,  $n$, as described by (\ref{1-10}). If we denote the frequency of emitted radiation as $\omega_0$, then
\begin{equation}
    \omega_0=M_{n+1}-M_n\simeq\frac{M_\text{P}}{2\sqrt{2n}}\simeq\frac{M_\text{P}^2}{4M}.
\end{equation}\label{3-1}
This agrees with the classical BH oscillation frequencies, which scale $1/M$.
We thus expect a BH to radiate with
a characteristic temperature $T\propto M_\text{P}^2/M$, matching the Hawking temperature.

The BH entropy can be expressed in terms of the following adiabatic invariant
\begin{equation}
    \label{H6}
    S_\text{BH}=8\pi\int_{M_\text{P}}^M\frac{dM}{\omega_0}=\frac{A}{4G}=4\pi M^2G,
\end{equation}
where $A=4\pi R_\text{S}^2$ is the BH horizon area, and $R_\text{S}=2MG$ is the Schwarzschild radius.
Because this spectrum is equally spaced, the possible values for the area of a massive BH are equally spaced.

One can obtain the quantum deformed (q-deformed) extension of the BH horizon area (\ref{3-12nona}) by replacing the ordinary ladder operators (\ref{new2}) with their q-extended ladder operators in which the Heisenberg--Weyl algebra turns into the quantum Heisenberg--Weyl algebra, $U_q
(h_4)$. The quantum Heisenberg--Weyl algebra, is the associative unital \cite{Chaichian1} $\mathbb C(q)$-algebra with generators
$\{a_+,a_-,q^{\pm N/2}\}$ with the following q-deformed commutation relations \cite{Chaichian1}
\begin{equation}\label{new4}
\begin{split}
&a_-a_+-q^{\frac{1}{2}}a_+a_-=q^{\frac{N}{2}},~~~[N,a_\pm]=\pm a_\pm,\\
&a_\pm^\dagger=a_\mp,~~N^\dagger=N,
\end{split}
\end{equation}
where $q$ is a primitive root of unity and is given by
\begin{equation}\label{3-14non}
q=\exp\left(\frac{2\pi i}{\mathcal N}\right),
\end{equation}
where $\mathcal N$ is the q-deformation parameter.  
The q-deformation is thought to be related to a fundamental dimensional constant. {Furthermore, the deformation parameter, $q$, must be a dimensionless function of such constant as well as of any specific system-characterizing features.}
The natural length scale of quantum gravity is the Planck length. 
{Thus, the} q-deformation parameter should be a function of the gravitational constant, $G$, or the Planck length squared \cite{Shabanov:1992yq} $\mathcal N=\mathcal N(L_\text{P}^2)$. In addition, one may expect that at the classical gravity limit, $L_\text{P}\rightarrow0$, $\mathcal N$ tends to infinity, and the theory tums into a classical gravity. This means that the q-deformation is a pure quantum gravity effect, and $\mathcal N\propto 1/L_\text{P}^2$. Regarding $\mathcal N$ being a dimensionless parameter, we need another length scale in which the ratio of the Planck length and the new length gives us the q-deformation parameter, i.e., $\mathcal N= L_q^2/L_\text{P}^2$. In section \ref{Cosmos}, we will show that this assumption leads us to an asymptotically de Sitter cosmological model, where $L_q$ 
%{will approach a de Sitter radius.}
plays the role of de Sitter radius. %However, this cosmological constant is 
{Moreover, we further take this new length scale as related to the ``cosmological'' constant, as consistently induced} by the q-deformation of the model. Hence, we consider $L_q \equiv \sqrt{3/\Lambda_q}$, where $\Lambda_q$ is q-cosmological constant, and $\mathcal N=L_q^2/L_\text{P}^2$ which leads to
\begin{equation}\label{3-14non1}
q=\exp\left(2\pi i\frac{L_\text{P}^2}{L_q^2}\right).
\end{equation}
Note that in our approach, $L_q$, similar to the Planck length, is a fundamental dimensional constant of quantum gravity theory \footnote{The CC, like the gravitational constant, is also used as a coupling constant in loop quantum gravity (LQG) and spinfoam frameworks \cite{Han:2010pz,Noui:2011im,Pranzetti:2014xva,Fairbairn:2010cp,Haggard:2015yda}. In LQG, a q-deformation has been derived as a mechanism to apply the theory's dynamics using $\Lambda$ and the deformation parameter, $q$, which is then given by (\ref{3-14non1}) \cite{Dupuis:2013lka}.
}. One can obtain the classical gravity limit of the theory by $L_q\rightarrow\infty$ or a vanishing q-cosmological constant.
 The above form of the deformation parameter realizes a holographic picture of quantum mechanics \cite{Jalalzadeh:2021oxi}.   Concretely, as shown in \cite{Jalalzadeh:2021oxi},  the Hilbert space of q-deformed neutral hydrogen gas in de Sitter space satisfies the strong holographic bound with the above deformation parameter.

The above quantum deformation of the BH gives us the following eigenvalues for the surface area and the mass of the BH \cite{Jalalzadeh:2022rxx}
\begin{equation}\label{3-19non}
A_n={4\pi L_\text{P}^2}\frac{\sin(\frac{\pi}{\mathcal{N}}(n+\frac{1}{2}))}{\sin(\frac{\pi}{2\mathcal{N}})},
\end{equation}
\begin{equation}\label{3-19non1}
M_n=\frac{M_\text{P}}{2}\sqrt{\frac{\sin(\frac{\pi}{\mathcal{N}}(n+\frac{1}{2}))}{\sin(\frac{\pi}{2\mathcal{N}})}},
\end{equation}
where $n=0,...,\mathcal{N}-1$.
Note that for $\mathcal{N}\rightarrow\infty$ (or equivalently, $\Lambda_q\rightarrow0$) the earlier eigenvalues will reduce to
(\ref{1-10}). Also, there is a two-fold degeneracy at the horizon's eigenvalues further at the BH's mass spectrum. 

To summarize the consequences of the above eigenvalue relations, for simplicity, let us consider $\mathfrak N$ is an odd natural number:

1) The area and the mass of the ground state $n=0$, as well as the state $n=\mathcal N-1$, are \cite{Jalalzadeh:2022rxx}
    \begin{equation}
  %  \begin{split}
        A_{0}=A_{\mathcal N-1}=4\pi L_\text{P}^2,\,\,\,\,
        M_{0}=M_{\mathcal N-1}=\frac{M_\text{P}}{2}. 
  %  \end{split}
    \end{equation}
    These show that the ground state's area and mass are not deformed, and their values are the same as the non-deformed spectrum obtained in (\ref{1-10}). Besides, the spectrum is bounded in which the most excited state, $n=\mathcal N-1$, has the same mass and area as the ground state. In addition, for $n\ll (\mathcal N-1)/2$ or $(\mathcal N-1)/2\ll n\leq\mathcal N-1$, Eqs. (\ref{3-19non}) and (\ref{3-19non1}) will reduce to the non-deformed spectrum obtained in (\ref{1-10}).

2) The above surface area spectrum of the
BH leads us to the following q-deformed entropy \cite{Jalalzadeh:2022rxx}
\begin{equation}
    \label{hoo1}
    S_q=\pi\frac{\sin(\frac{\pi}{\mathcal{N}}(n+\frac{1}{2}))}{\sin(\frac{\pi}{2\mathcal{N}})}=\pi\frac{\sin(\frac{\pi R_S^2}{2G\mathcal N})}{\sin(\frac{\pi}{2\mathcal N})},~~~~\frac{\pi R_S^2}{2G\mathcal N}\leq\frac{\pi}{2},
\end{equation}
where $R_S$ is the Schwarzschild radius of the BH.

With keeping one eye on Eq. (\ref{new2}), the commutation relation of the q-deformed $x$ and $p$ is
\begin{equation}\label{shhh}
    [x,p]|n\rangle=%i\frac{\cos\left(\frac{\pi}{\mathcal{N}}(n+\frac{1}{2}) \right)}{\cos(\frac{\pi}{2\mathcal N})}|n\rangle=
    i\frac{\cos\left(\frac{\pi G\Lambda_q}{3}(n+\frac{1}{2}) \right)}{\cos(\frac{\pi G\Lambda_q}{6})}|n\rangle.
\end{equation} 
According to Eq. (\ref{hoo1}), the q-entropy of a BH reaches its greatest value for $n=(\mathcal N-1)/2$. In this case, regarding the second equality in Eq. (\ref{hoo1}), the Schwarzschild horizon radius is $R_S^2=G\mathcal{N}=L_q^2$. On the other hand, Eq. (\ref{shhh}) shows that for $n=(\mathcal N-1)/2$, (or when the BH radius equals the {de Sitter} radius) $x$ and $p$ commute. As a result, the classical state is the one with the most entropy or maximum radius. Note that the effective Schwarzschild horizon radius is different from $R_S$ and is given by
\begin{equation}
    R_\text{eff}=L_\text{P}\sqrt{\frac{\sin(\frac{\pi R_S^2}{2\mathcal{N}L_\text{P}^2})}{\sin(\frac{\pi}{2\mathcal{N}})}}.
\end{equation}

\section{q-deformed Friedmann and Raychaudhuri equations}\label{FR}

Our starting point is the Friedmann--Lemaître--Robertson--Walker (FLRW)  line element in comoving coordinates
\begin{equation}
    \label{Sam1}
ds^2=h_{ab}dx^adx^b+R^2d\Omega^2_{(II)},~~~a=0,1,
\end{equation}
where $d\Omega^2_{(II)}$ is the line element of the standard 2-sphere, $R(r,t)=ra(t)$ is the areal radius, and $h_{ab}=\text{diag}{(-1,\frac{a(t)^2}{1+kr^2})}$. {Moreover}, $x^0=t$, $x^1=r$, and as usual, the open, flat, and closed universes correspond to
$k = -1, 0, 1$, respectively

Several authors \cite{Bak:1999hd,Bousso:2004tv,Collins:1992eca,GalvezGhersi:2011tx,Hayward:1997jp,Hayward:1998ee} have 
{taken}
%adjusted 
the de Sitter event (and apparent) horizon thermodynamic formulae to 
{encompass, constituting a larger, upper bound regarding}
%account for 
a generic FLRW space's non-static apparent horizon, 
which differs from the event horizon. 
{For the sake of clarity, let us emphasize that it}
 is widely proposed in dynamical spacetimes that the apparent horizon 
 {represents}
 %reflects 
 a causal horizon {associated} 
 %linked 
 to gravitational {temperature and entropy (and surface gravity)}. If this is correct, the same might be said about cosmic horizons. It was suggested in Refs. \cite{Bousso:2004tv,Davies:1988dk,Frolov:2002va,GalvezGhersi:2011tx} {that the} FLRW {thermodynamical} 
 %space 
 event horizon 
 %thermodynamics 
 needs to be better defined (except for 
 {the case of} de Sitter space). The FLRW apparent horizon's Hawking radiation was calculated by the authors of \cite{Jiang:2009kzr,Zhu:2008hn}. It was rederived in references \cite{Cai:2008gw,Medved:2002zj} by applying the Hamilton--Jacobi method \cite{Angheben:2005rm,Nielsen:2005af,Visser:2001kq} to the Parikh--Wilczek approach, which was initially developed for BH horizons \cite{Parikh:1999mf}.

A substantial amount of research has been conducted on the 
{bridging of geometry and} thermodynamics 
{within} 
%of 
FLRW spaces (see, for example, \cite{Akbar:2006kj,Cai:2005ra} and references therein). The thermodynamical properties of the FLRW apparent horizon are described in Ref. \cite{DiCriscienzo:2009kun}. {Allow us to point to the}
%There are also 
Kodama vector, Kodama--Hayward surface gravity, and 
{corresponding} Hawking temperature calculations. The Kodama--Hayward temperature of the FLRW apparent horizon is given by
\begin{equation}
    \label{Temp}
    T_\text{AH}=\frac{\kappa_\text{kodama}}{2\pi}=-\frac{1}{2\pi R_\text{AH}}\left(1-\frac{\dot R_\text{AH}}{2HR_\text{AH}} \right),
\end{equation}
where $H=\dot a/a$ is the Hubble parameter, $\kappa_\text{Kodama}$ is the Kodama surface gravity of the apparent horizon, and $R_\text{AH}$ is the radius of the apparent horizon given by
\begin{equation}
    \label{apparent}
    R_\text{AH}=\frac{1}{\sqrt{H^2+\frac{k}{a^2}}}.
\end{equation}

Let us assume that the FLRW universe is sourced by a perfect fluid with energy-momentum tensor
\begin{equation}
    \label{energy-momentum}
    T_{\mu\nu}=(\rho+p)u_\mu u_\nu+g_{\mu\nu}p,
\end{equation}
where $\rho$, $p$, and $u_\mu$ are the fluid's total energy density, pressure, and 4-velocity field, respectively. The perfect fluid could be a combination of non-interacting dust (cold dark and baryonic matters), $\rho_\text{c}$, and radiation, $\rho_\text{rad}$ \footnote{We use a subscript $x$ as one of ``c'' for the dust of baryons plus dark matter, ``rad'' for radiation (photons plus relativistic neutrinos), ``m'' for baryons plus dark matter plus radiation, and ``DE'' for the effective dark energy.}
\begin{equation}
    \begin{split}
       & \rho=\rho_\text{c}+\rho_\text{rad},\\
    &   p=p_\text{rad},~~~~p_\text{c}=0,
  \end{split}
\end{equation}

Regarding these considerations, the unified first law of thermodynamics on the apparent horizon \cite{Hayward:1997jp,Hayward:1998ee} is given by
\begin{equation}
    \label{Firstlaw}
    T_\text{AH}\dot S_\text{AH}=\dot M_\text{AH}+\frac{p-\rho}{2}\dot V_\text{AH},
\end{equation}
where an overdot means time derivative, $S_\text{AH}$ is the entropy of the apparent horizon, and
\begin{equation}
    \label{mass}
    V_\text{AH}=\frac{4\pi}{3}R_\text{AH}^3,~~~
    M_\text{AH}=\rho V_\text{AH},
\end{equation}
are the areal volume and the Misner--Sharp--Hernandez mass contained inside the apparent horizon, respectively.

Inserting the Kodama--Hayward temperature (\ref{Temp}), the areal volume, and  the Misner--Sharp--Hernandez mass defined in (\ref{mass}) into the unified first law of thermodynamics (\ref{Firstlaw}) gives us
\begin{equation}
    \label{Firstlaw2}
    \dot S_\text{AH}=8\pi^2HR_\text{AH}^4H(\rho+p).
\end{equation}
If we postulate that the covariant conservation equation is satisfied by the energy-momentum tensor of the universe's perfect fluid composition (\ref{energy-momentum}), then we find 
\begin{equation}
    \label{Contin}
    H\left(\rho_{i}+p_{i}\right)=-\frac{\dot\rho_{i}}{3},~~~~i=\text{c}, \text{rad}.
\end{equation}
Inserting this relation in the r.h.s of Eq. (\ref{Firstlaw2}) simplifies it into
\begin{equation}
    \label{Firstlaw3}
    \frac{1}{8\pi^2R_\text{AH}^4}\dot S_\text{AH}=-\frac{1}{3}\dot\rho.
\end{equation}

Now, our assumption is to take the entropy associated with the apparent horizon in the form of q-deformed entropy 
(\ref{hoo1}) and replace the BH horizon radius, $R_S$, with the apparent horizon radius $R_\text{AH}$. This gives us the q-deformed entropy of the apparent horizon
\begin{equation}
    \label{Qentropy}
    S_q=\pi\frac{\sin(\frac{\gamma R_\text{AH}^2}{G})}{\sin(\gamma)},~~~~~0\leq \frac{\gamma R_\text{AH}^2}{G}\leq\frac{\pi}{2},
\end{equation}
where $\gamma=\pi/(2\mathcal N)=\frac{\pi}{2}\left(\frac{L_\text{P}}{L_q} \right)^2=\frac{\pi}{6}G\Lambda_q$. Let us first discuss some of {the} 
%real 
outcomes of this {expression for the} entropy before delving {further}.
%into the cosmological context of it. 

\begin{enumerate}
    \item Eq. (\ref{Qentropy}) is non-classical by definition. According to the noncommutative perspective of quantization \cite{Jalalzadeh:2022rxx,Jalalzadeh:2017jdo,Majid:2008iz,Majid:1999tc,Majid:2014afa,Papageorgiou:2010ud},
     $\hbar$ and $\Lambda_q$ are both quantization parameters, and the classical limit of the model can be realized by establishing the $\hbar\rightarrow0$ limit. On the other hand, $\Lambda_q\rightarrow0$ gives us quantum gravity without a q-cosmological constant. 

\item Our cosmological model did not account for a traditional CC, but our {subsequently implemented}  quantum deformation led to the emergence of an effective one, strongly linked to the natural number $\mathcal{N}$ defined by (\ref{3-14non}). This constant directly results from the {assumption of a} finite number of states in Hilbert spaces \cite{Jalalzadeh:2017jdo}.
 
  \item  Regarding our discussion on the entropy of q-deformed BHs in 
  the previous section, the 
  {interval}
  %hole distance 
  $0\leq \frac{\gamma R_\text{AH}^2}{G}\leq{\pi}$ is divided to $0\leq \frac{\gamma R_\text{AH}^2}{G}\leq\frac{\pi}{2}$, and $\frac{\pi}{2}\leq \frac{\gamma R_\text{AH}^2}{G}\leq{\pi}$. As the first interval, $0\leq \frac{\gamma R_\text{AH}^2}{G}\leq\frac{\pi}{2}$ represents an expanding universe, the second interval realizes a contracting universe. These universes coexist simultaneously at the q-deformed quantum cosmology level and have an entangled quantum state \cite{Jalalzadeh:2022rxx}. By 
  observation, the state collapses it into expanding or contracting universe. Here, we assumed
  (regarding the cosmological observations), 
  the state has collapsed into the expanding universe.
  
  \item   The apparent horizon radius, $R_\text{AH}$, is quantized according to
    Eq. (\ref{hoo1}). As a result, $R_\text{AH}=L_\text{P}$ gives the smallest value of the apparent horizon radius, whereas $L_q$ is the larger scale within our knowledge of physics models and to ensure a finite Hilbert space. However, because we are interested in the late-time evolution of the universe, we assumed the apparent horizon radius was zero at the Big Bang for simplicity.
\end{enumerate}

Implying the q-deformed entropy (\ref{Qentropy}) for FLRW spaces 
into (\ref{Firstlaw3}), gives us
\begin{equation}
    \label{Firstlaw4}
    \cos(\frac{\gamma}{G}R_\text{AH}^2)\frac{\dot R_\text{AH}}{R_\text{AH}^3}=-\frac{4\pi G}{3}\frac{\sin(\gamma)}{\gamma}\dot \rho.
\end{equation}
Integration of the above differential equation yields
\begin{multline}
    \label{Freed}
    \frac{\cos(\frac{\gamma}{G}R_\text{AH}^2)}{R_\text{AH}^2}+\frac{\gamma}{G}\left\{ \Si\left(\frac{\gamma}{G}R_\text{AH}^2\right)-\Si\left(\frac{\pi}{2}\right)\right\}=\\\frac{\sin(\gamma)}{\gamma}\frac{8\pi G}{3}\rho,
\end{multline}
where 
\begin{equation}
    \Si(x)=\int_0^x\frac{\sin(y)}{y}dy,
\end{equation}
is the integral sine function. Note that, regarding $0\leq \frac{\gamma}{G}R_\text{AH}^2\leq\pi/2$, the constant of integration is chosen in such a way that the above equation becomes trivial for $\frac{\gamma}{G}R_\text{AH}^2=\frac{\pi}{2}$. This means that, as we expected, the FLRW space {would be instead}
%is 
asymptotic de Sitter, i.e., $ 1/R_\text{AH}^2=\frac{\Lambda_q}{3}$.

One can simplify Eq. (\ref{Freed}) by considering the value of $\gamma=\frac{\pi}{2}\left(\frac{L_\text{P}}{L_q} \right)^2\simeq10^{-123}$. This shows that $\sin(\gamma)/\gamma\simeq1$. Hence, (\ref{Freed}) reduces to
\begin{multline}
    \label{Freed1}
    \frac{\cos(\frac{\pi\Lambda_q}{6}R_\text{AH}^2)}{R_\text{AH}^2}=\frac{8\pi G}{3}\rho\\+\frac{\pi\Lambda_q}{6}\left\{ \Si\left(\frac{\pi}{2}\right)-\Si\left(\frac{\pi\Lambda_q}{6}R_\text{AH}^2\right)\right\}.
\end{multline}
Equation (\ref{Freed1}) is the q-deformed Friedmann equation based on the q-deformed entropy. Therefore, {assuming} 
%initiating with
the unified first law of thermodynamics 
{applied} at the apparent horizon of an FLRW universe, plus supposing that the apparent horizon area includes quantum deformed characteristics, we derive the corresponding modified Friedmann equation of an FLRW universe with any spatial curvature.

The Raychaudhuri equation, often known as the second Friedmann equation, may be obtained
directly using Eq. (\ref{Firstlaw2}) and the Friedmann equation (\ref{Freed1}). Inserting the q-entropy (\ref{Qentropy}) into (\ref{Firstlaw2}) and using 
\begin{equation}
    \dot R_\text{AH}=-R_\text{AH}^3\left(\frac{\ddot a}{a}-\frac{1}{R_\text{AH}^2}\right)H,
\end{equation}
we obtain
\begin{equation}
    -\frac{\ddot a}{a}+\frac{1}{R_\text{AH}^2}=\frac{4\pi G}{\cos(\frac{\gamma}{G}R_\text{AH}^2)}\frac{\sin(\gamma)}{\gamma}(\rho+p).
\end{equation}
This equation, combined with the Friedmann equation (\ref{Freed1}), gives the q-deformed Raychaudhuri equation
\begin{multline}
    \label{Raych}
    \cos(\frac{\pi\Lambda_q}{6}R_\text{AH}^2)\frac{\ddot a}{a}=-\frac{4\pi G}{3}(\rho+3p)\\+\frac{\pi\Lambda_q}{6}\left\{ \Si\left(\frac{\pi}{2}\right)-\Si\left(\frac{\pi\Lambda_q}{6}R_\text{AH}^2\right)\right\}.
\end{multline}

Note that at the limit classical limit of quantum geometry, ${\Lambda_q}\rightarrow0$ (or $\mathcal N\rightarrow\infty$), Eqs. (\ref{Freed1}) and (\ref{Raych}) reduce to the standard  Friedmann and Raychaudhuri equations
\begin{equation}
    \begin{split}
        &H^2+\frac{k}{a^2}=\frac{8\pi G}{3}\rho,\\
       & \frac{\ddot a}{a}=-\frac{4\pi G}{3}(\rho+3p),\\
       &\rho=\rho_\text{c}+\rho_\text{rad}, ~~~p=p_\text{rad},
    \end{split}
\end{equation}
where {$\rho_\text{c}$ is the energy density of cold matter}, and $\rho_\text{rad}$, and $p_\text{rad}$ are the energy density and the pressure of the radiation. 

\section{q-deformed cosmology}\label{Cosmos}

In the remainder of this paper, for convenience, we focus on the spatially flat universe, namely $k=0$. Then, $1/R_\text{AH}=H^2$, and one can rewrite the q-deformed  Friedmann and Raychaudhuri equations (\ref{Freed1}) and (\ref{Raych}) as
\begin{equation}\label{QF1}
    H^2=\frac{8\pi G}{3}(\rho+\rho_\text{DE}),
\end{equation}
\begin{equation}\label{QF2}
    \dot H=-{4\pi G}(\rho+p+\rho_\text{DE}+p_\text{DE}),
\end{equation}
where we have introduced an effective q-dark energy sector, with energy density, $\rho_\text{DE}$, and pressure, $p_\text{DE}$, respectively, {with}
%of 
the following form
\begin{multline}
    \label{Qrho}
    \rho_\text{DE}=\frac{3}{4\pi G}\sin^2\left(\frac{\pi{\Lambda_q}}{12H^2}\right)H^2+\\\frac{\Lambda_q}{16G}\left\{\Si\left(\frac{\pi}{2}\right)-\Si\left(\frac{\pi\Lambda_q}{6H^2}\right)\right\},
\end{multline}
\begin{multline}
    \label{QP}
    p_\text{DE}=-\frac{1}{4\pi G}\sin^2\left(\frac{\pi\Lambda_q}{12H^2}\right)\left[2\dot H+3H\right]\\-\frac{\Lambda_q}{16G}\left[ \Si\left(\frac{\pi}{2}\right)-\Si\left(\frac{\pi\Lambda_q}{6H^2}\right) \right].
\end{multline}
These give us the EoS parameter for the q-dark energy sector
\begin{equation}
    \label{QO}
    \omega_\text{DE}=-1-\frac{8\sin^2\left(\frac{\pi\Lambda_q}{12H^2}\right)\dot H}{12\sin^2\left(\frac{\pi\Lambda_q}{12H^2}\right)H^2+\Lambda_q\pi\left[ \Si\left(\frac{\pi}{2}\right)-\Si\left(\frac{\pi\Lambda_q}{6H^2}\right) \right]}.
\end{equation}
The equations are more easily expressed in terms of dimensionless variables. Defining the density parameters of the {matter} (radiation plus cold dark and baryonic matters), the q-deformation parameter, $\Lambda_q$, and effective q-dark energy
\begin{equation}
    \label{Omegas}
    \Omega^\text{(m)}=\frac{8\pi G\rho_\text{m}}{3H^2},~~~ \Omega^{(q)}=\frac{\Lambda_q}{3H^2},~~~ \Omega^\text{(DE)}=
    \frac{8\pi G\rho_\text{DE}}{3H^2},
\end{equation}
then, the continuity  equation (\ref{Contin}) and 
q-deformed Raychaudhuri equations are expressed as
\begin{equation}\label{Contin2}
   \frac{\cos(\frac{\pi\Omega^{(q)}}{2})}{3\Omega^\text{(i)}}\frac{d\Omega^{(i)}}{d\tilde N}=\sum_j(1+\omega_j)\Omega^\text{(j)}-(1+\omega_i),~~~i=\text{c},\text{rad},
\end{equation}
and
\begin{equation}
    \label{Rich22}
\frac{\cos(\frac{\pi\Omega^{(q)}}{2})}{H}\frac{dH}{d\tilde N}=-\frac{3}{2}\sum_j(1+\omega_j)\Omega^\text{(j)},
\end{equation}
where $\tilde N=\ln(a/a_0)$ is the e-folding factor. The equivalent expression of the above equation can be obtained by using the definition of the q-cosmological constant's density parameter
\begin{equation}
    \label{Rich23}
    \frac{\cos(\frac{\pi\Omega^{(q)}}{2})}{\Omega^{(q)}}\frac{d\Omega^{(q)}}{d\tilde N}=3\sum_j(1+\omega_j)\Omega^\text{(j)}.
\end{equation}

In addition, the q-deformed  Friedmann (\ref{QF1}) takes the following form
\begin{multline}
    \label{QF33}
\Omega^\text{(DE)}=1-\Omega^\text{(m)}=1-\cos(\frac{\pi\Omega^{(q)}}{2})+\\\frac{\pi\Omega^{(q)}}{2}\left\{\Si\left(\frac{\pi}{2}\right)-\Si\left(\frac{\pi\Omega^{(q)}}{2}\right)\right\},~~~i=\text{c},\text{rad}.
\end{multline}
This allows us to eliminate $\Omega^{(q)}$ (and $\Omega^{(DE)}$) in terms of $\Omega^\text{(m)}$. 

Eqs. (\ref{Contin2}) and (\ref{Rich22}) allow us to calculate the density parameters of the radiation, the cold matter, and the q-deformation  
\begin{equation}
  \begin{split}
& \Omega^\text{(rad)}= \Omega^\text{(rad)}_0\left(\frac{H_0}{H}\right)^2(1+z)^4,\\
 &     \Omega^\text{(c)}= \Omega^\text{(c)}_0\left(\frac{H_0}{H}\right)^2(1+z)^3,\\
 &\Omega^{(q)}=\Omega^{(q)}_0\left(\frac{H_0}{H}\right)^2,
  \end{split}
\end{equation}
where $z=a_0/a-1$ is the redshift, and also $ \Omega^\text{(rad)}_0$, $ \Omega^\text{(c)}_0$, and $\Omega^{(q)}_0$ are the density parameters of the radiation, the cold matter, and q-deformation parameter at the present epoch. Inserting these into the Hamiltonian constraint (\ref{QF33}) gives us
\begin{multline}
    \label{H123}
\cos(\frac{\pi\Omega^{(q)}_0}{2E^2})E^2=\frac{\pi\Omega^{(q)}_0}{2}\Big\{\Si\left(\frac{\pi}{2}\right)-\\\Si\left(\frac{\pi\Omega^{(q)}_0}{2E^2}\right)\Big\}
+\Omega^\text{(rad)}_0(1+z)^4+\Omega^\text{(c)}_0(1+z)^3,
\end{multline}
where $E={\frac{H}{H_0}}$.

\begin{figure}[ht]
\centering
\includegraphics[width=8cm]{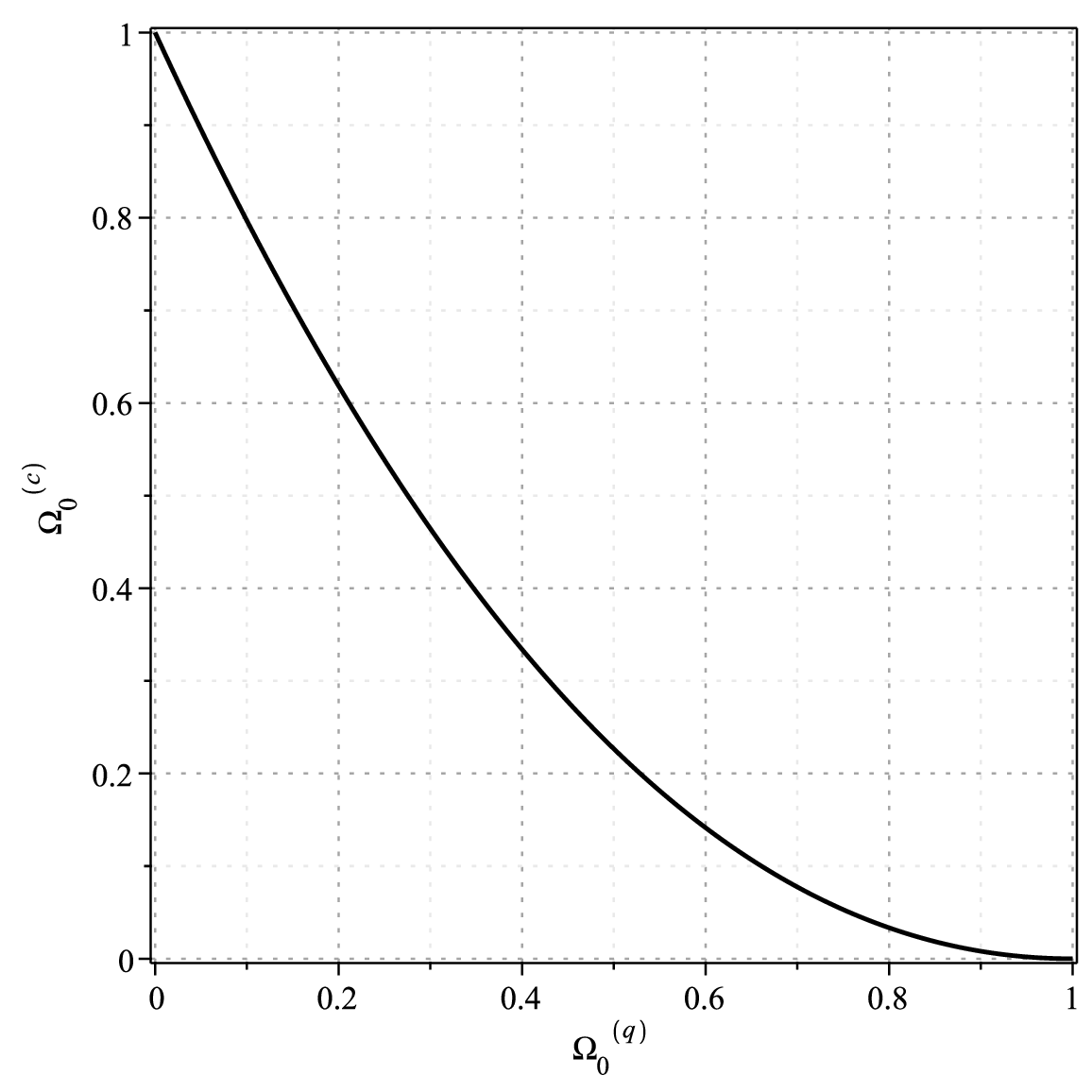}
\caption{Plot of the density parameter of the matter (radiation, plus cold matter) versus the density parameter of the q-cosmological constant. We plot this figure using Eq. (\ref{H123}). }\label{ppp1}
\centering
\end{figure}

Evaluating (\ref{H123}) at the present epoch gives us the relation between the density parameters, which we now write as
\begin{multline}
    \label{H123aa}
    \cos(\frac{\pi\Omega^{(q)}_0}{2})-\frac{\pi\Omega^{(q)}_0}{2}\Big\{\Si\left(\frac{\pi}{2}\right)-\\\Si\left(\frac{\pi\Omega^{(q)}_0}{2}\right)\Big\}=\Omega^\text{(m)}_0.
\end{multline}
It would be beneficial to delve deeper into this equation.
Fig. \ref{ppp1} illustrates the profound influence of the q-deformation parameter on our universe. A significant value of $\Omega^{(q)}_0$ (which implies a smaller current Hubble distance of $L_q$) decreases the relevance of the baryonic and dark matters. Conversely, a q-deformation parameter of zero elevates cold matter to the position of the primary driving force behind the Universe's evolution. Consequently, very small q-deformation can significantly impact cosmology during later stages.

%%%%%%%%%%%%%%%%%%%%%
%%%%%%%%%%%%%%%%%

\begin{figure}[ht]
\centering
\includegraphics[width=8cm]{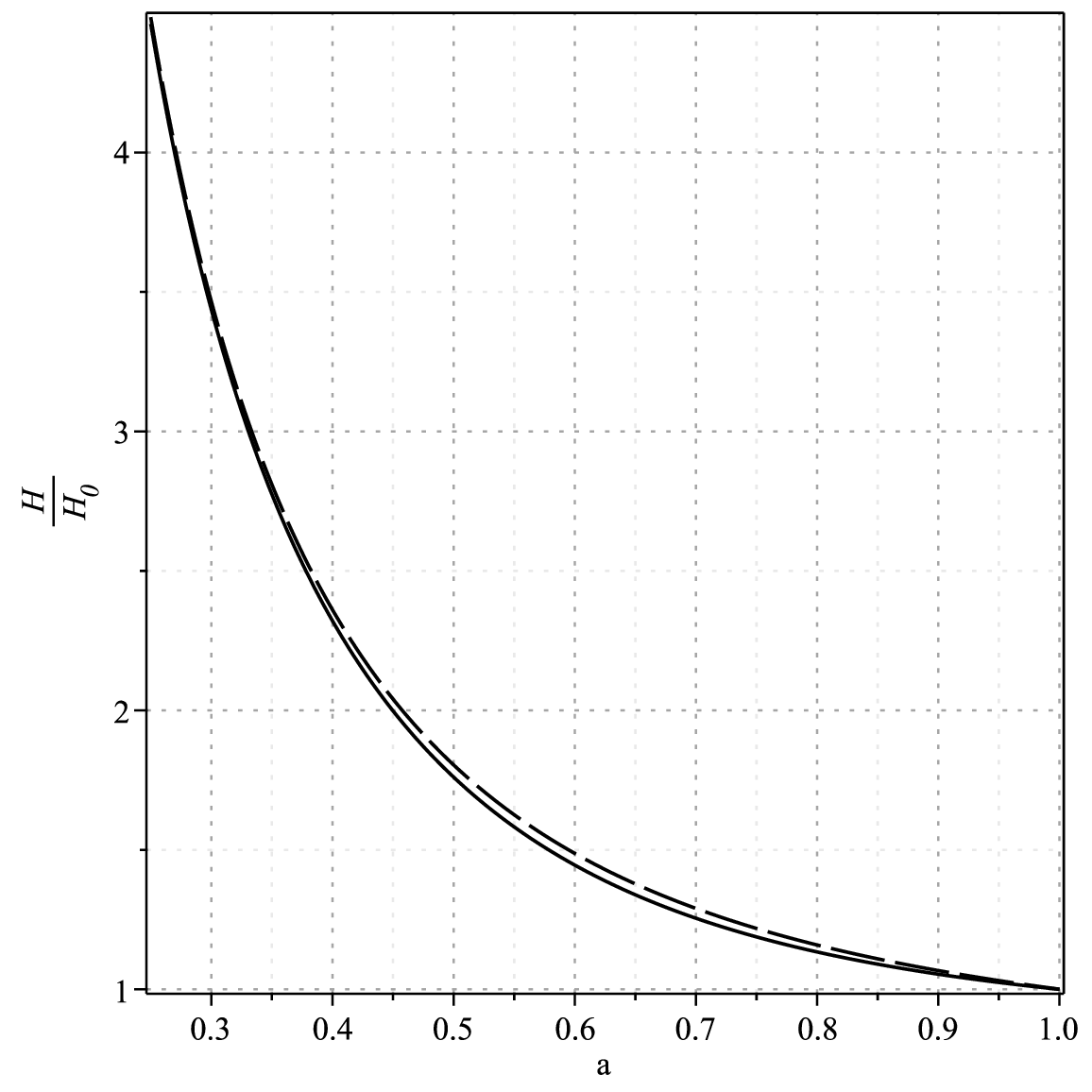}
\caption{{Plot of $H/H_0$ (radiation$+$cold matter) versus the scale factor in flat $\Lambda$CDM and q-deformed cosmology. The black line (dash) shows the evolution of the Hubble parameter in q-deformed ($\Lambda$CDM) cosmology. We plot these figures using Eq. (\ref{H123}) for q-deformed cosmology and the standard equation of $H/H_0$ in the standard model of cosmology, respectively. We used $\Omega^{(q)}_0=0.43$, $\Omega^\text{(c)}_0=0.3$ in q-cosmology and $\Omega^{(\Lambda)}_0=0.4$ in $\Lambda$CDM models. }}\label{ppp12}
\centering
\end{figure}

It is clear that the above expression is not analytically solvable. Hence, we use the global series expansion of $\cos(x)+x\Si(x)=\sum_{n=0}^{\infty}\frac{(-1)^{n+1}x^{2n}}{(2n)!(2n-1)}$ to simplify Eq. (\ref{H123}) and obtain $\Omega^{(q)}_0$ in terms of density parameter of the matte at the present epoch. Using this series expansion, Eq. (\ref{H123aa}) up to $\mathcal{O}((\Omega^{(q)})^4)$ simplifies to
\begin{equation}
    1+\frac{1}{2}\left(\frac{\pi \Omega^{(q)}_0}{2}\right)^2-\Si(\frac{\pi}{2})\left(\frac{\pi \Omega^{(q)}_0}{2}\right)-\Omega^\text{(m)}_0=0.
\end{equation}
{It is essential to note that we have not used any approximations until this point. However, the q-deformed Friedmann equation (\ref{QF1})  and Raychaudhuri equation (\ref{QF2}) do not have analytical solutions, making it complicated to calculate different quantities, like the age of the universe or distance modulus. Although numerical methods can be used to perform these calculations without approximation, our primary goal in this paper is to enable readers to follow the calculations easily. In the early universe and during the radiation-dominated epoch, Eqs. (\ref{QF1}) and (\ref{QF2}) are equivalent to the Friedmann equation and Raychaudhuri equations in the $\Lambda$CDM model with exceptional approximation; please see Fig. \ref{ppp12}. Nonetheless, the differences between the standard model and our model are only noticeable at small redshifts. }
%%%%%%%%%%%%%%%%%%%%%
%%%%%%%%%%%%%%%%%
%The root of this equation gives us
%\begin{equation}\label{M1}
  %  \Omega^{(q)}_0=\frac{2}{\pi}\left\{\Si(\frac{\pi}{2})-\sqrt{\Si^2(\frac{\pi}{2})-2(1-\Omega^\text{(m)}_0)} \right\}.
%\end{equation}
Applying the above procedure for Eq. (\ref{H123}) results\footnote{It should be noted that the approximation (\ref{H124}) is inappropriate for negative redshifts, and therefore the higher order of approximation must be used.  }
\begin{multline}\label{H124}
    \Omega^{(q)}=2 \Omega^{(q)}_0\Bigg\{ \beta\Omega^{(q)}_0+f(z)+\\\sqrt{\left(\beta \Omega^{(q)}_0+f(z) \right)^2-4\left(\beta\Omega^{(q)}_0+\Omega^\text{(m)}_0-1 \right)} \Bigg\}^{-1},
\end{multline}
where $\beta=\frac{\pi}{2}\Si(\frac{\pi}{2})$, and
\begin{equation}
    f(z)=\sum_j\Omega^\text{(j)}_0(1+z)^{3(\omega_j+1)}, ~~~i=\text{c},\text{rad}.
\end{equation}

{Also, in the first approximation, one can simplify Eq. (\ref{H123}) to find an explicit expression of the Hubble parameter
\begin{multline}
    \label{Hubble}
    \left(\frac{H}{H_0}\right)^2=\Big\{f(z)+\\\frac{\pi}{2}\Si(\frac{\pi}{2})\Omega^{(q)}_0\Big\}\left\{1-\frac{\pi^2\left(\Omega_0^{(q)}\right)^2}{8\left(f(z)+\frac{\pi}{2}\Si(\frac{\pi}{2})\Omega^{(q)}_0\right)^2} \right\},
\end{multline}
where the higher-order terms are omitted.
This equation is tantamount to the Friedmann equation in the standard flat $\Lambda$CDM cosmology, supplemented with a correction term articulated in the last term. In addition, the induced effective density parameter of the CC at the present epoch is $\frac{\pi}{2}\Si(\frac{\pi}{2})\Omega^{(q)}_0$. }

{To analyze cosmological models, datasets obtained from SNIa observations are particularly valuable as they serve as primary evidence for the Universe's accelerated expansion.
To achieve optimal outcomes with SNIa data, we commence with the observed distance modulus produced by SNIa detections and compare it against the theoretical value. For this study, we utilize the Pantheon sample, an up-to-date SNIa dataset that encompasses 1048 distance modulus $\mu$ at various redshifts within the $0.01 < z < 2.26$ range \cite{Pan-STARRS1:2017jku}.
%Therefore, we utilized a sample of 1048 distance modulus datasets from Union 2.1 in combination with SNIa \cite{2012ApJ85S}. In order to study the evolution of the Universe, the redshift-luminosity distance relation is a prominent observational tool. As a result of the Universe's expansion, light emitted from a distant luminous object becomes redshifted, and an equation for luminosity distance in terms of redshift can be obtained.
By utilizing luminosity distance, one can determine the distance modulus, which is presented as
\begin{equation}\label{modul}
    \mu(z)=25+5\log_{10}\left((1+z)\int_0^z\frac{dz'}{H(z')} \right).
\end{equation}
}
\begin{figure}[ht]
\centering
\includegraphics[width=8cm]{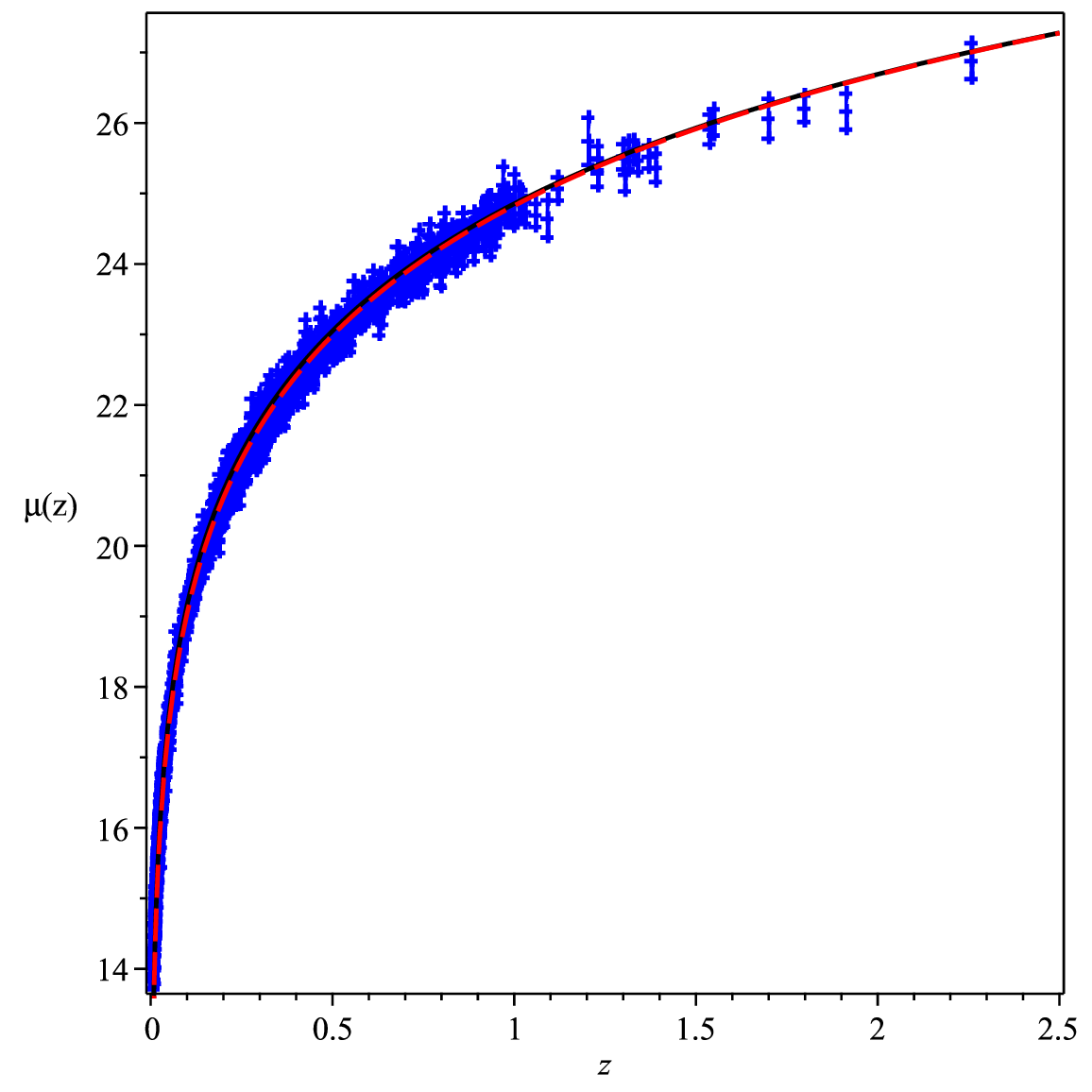}
\caption{Evolution of Luminosity Distance, $\mu(z)$, versus redshift $z$. The black line denotes the best fit for the redshift evolution of the distance modulus obtained from Eqs. (\ref{modul}) and (\ref{H123}). We used $\Omega^{(q)}_0=0.43$, $\Omega^\text{(c)}_0=0.3$ and $H_0=67.27~\text{Km}\cdot \text{sec}^{-1}\cdot\text{Mpc}^{-1}$, and the absolute B-band magnitude of a fiducial SNIa is $M_0=-19.5$ \cite{Wang:1999bz}.
 The red line (dots) denotes the redshift evolution of the distance modulus in the standard flat $\Lambda$CDM model with  $\Omega_0^{(\Lambda)}=0.7$. The distance modulus measurements we use are taken from
 Ref. \cite{Pan-STARRS1:2017jku}.}\label{modulus}
\centering
\end{figure}

Cosmological observations show that the energy density parameter of the radiation is negligible at the present epoch, $\Omega^\text{(rad)}_0\simeq8.7\times 10^{-5}$. Hence, for simplicity, we assume that the matter content of the model is only cold matter. Thus, one can safely ignore the radiation density parameter in the field equations, i.e., $\Omega^\text{(m)}=\Omega^\text{(c)}$. As a result of this assumption, $f(z)$ simplifies into
\begin{equation}
     f(z)=\Omega^\text{(c)}_0(1+z)^3.
\end{equation} 
{Figure \ref{modulus} 
displays the Pantheon Survey as the standard Hubble diagram of SNIa (absolute
magnitude $M_0=-19.5$).
 By utilizing this dataset, we are able to determine the optimal values for the density parameter of cold matter $\Omega^\text{(c)}_0$ and  $ \Omega^{(q)}_0$
\begin{equation}\label{Fitt}
    \Omega^{(q)}_0=0.43,~~~ \Omega^\text{(c)}_0=0.3,
\end{equation}
where the value of $H_0$ is set to be 67.27 (Km/s)/Mpc based on the Planck 2018 results \cite{Planck:2018vyg}.
Furthermore, in this figure, our model has been juxtaposed against the standard flat cosmology. It is evident from the figure that the two models display a remarkable degree of conformity with each other. }

In addition, using the approximations $\Omega^\text{(m)}=\Omega^\text{(c)}$, the energy density parameter of q-dark energy and its EoS parameter, defined by Eqs. (\ref{Omegas}) and (\ref{QO}) simplify 
\begin{equation}\label{M2}
  \begin{split}
        \Omega^\text{(DE)}&=1-\frac{\Omega^{(q)}}{\Omega^{(q)}_0}f(z),\\
        \omega_\text{DE}&=-1+\frac{\pi\left(\Omega^{(q)}\right)^2f(z)}{4\Si(\frac{\pi}{2})\Omega^{(q)}_0}\left\{1+\frac{\pi\Omega^{(q)}}{4\Si(\frac{\pi}{2})}\right\}.
  \end{split}
\end{equation}
{By substituting the values of $\Omega^\text{(c)}_0$ and $\Omega^{(q)}_0$ into the above equations, we find the energy density parameter and EoS parameter of q-dark energy at the present epoch
\begin{equation}
 \Omega^\text{(DE)}_0=0.7,~~~\omega_\text{DE0}=-0.91.
\end{equation}}

\begin{figure}[ht]
\centering
\includegraphics[width=8cm]{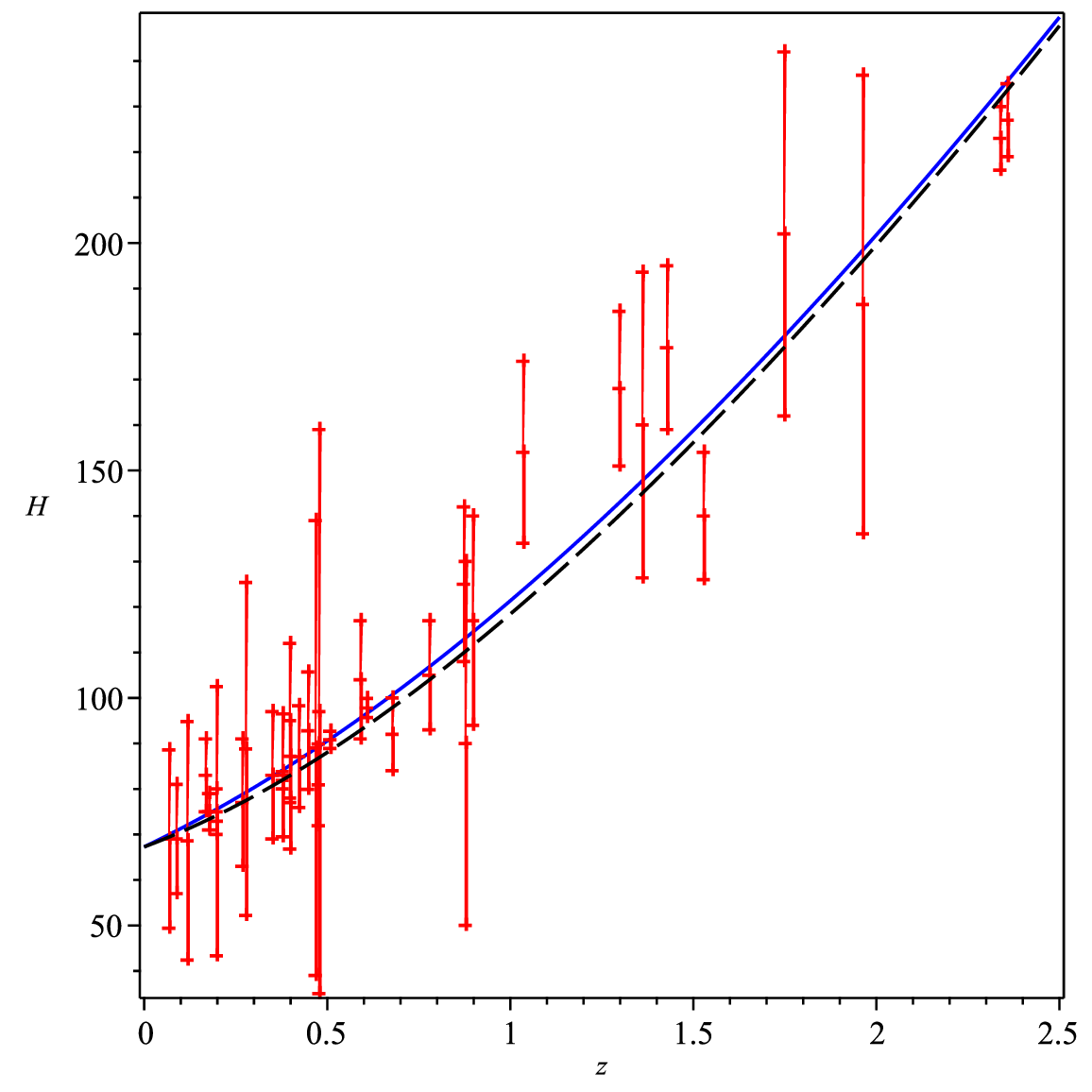}
\caption{The evolution of $H(z)$ (in units $\text{Km}\cdot \text{sec}^{-1}\cdot\text{Mpc}^{-1}$) versus redshift $z$ with error bars. The blue line denotes the dynamics of the Hubble parameter obtained from Eq. (\ref{H123}).
The values of  $\Omega^{(q)}_0$, $\Omega^\text{(c)}_0$ and $H_0$ are same as the Eq. \ref{modul}, and we assumed (according to the standard flat $\Lambda$CDM model)  $\Omega_0^{(\Lambda)}=0.7$.
 The dash-line denotes the evolution of the Hubble parameter in the standard flat $\Lambda$CDM model with the same values of $H_0$, $\Omega^\text{(c)}_0$ and $\Omega_0^{(\Lambda)}$. The Hubble parameter measurements we use are taken from
Table 1 of Ref. \cite{Yu:2017iju}. }\label{hubble1}
\centering
\end{figure}
{Furthermore, we were able to achieve a satisfactory level of agreement for the Hubble parameter by utilizing the fitting parameters (\ref{Fitt}). By referring to the data presented in Table 1 of \cite{Yu:2017iju}, we plotted the Hubble parameter $H$ versus the redshift $z$ in Fig. \ref{hubble1}. Our model demonstrated a high level of consistency with the standard model of cosmology, as evidenced by the figure. It is worth noting that all entropic-force models tend to fit the modulus distance data and Hubble parameter points in a similar fashion (cf. \cite{Komatsu:2014lsa,Tartaglia:2008nu,Jana:2014aca,Zamora:2022cqz,Sharma:2021zjx}). While more comprehensive data analysis (e.g., using a covariance matrix) is necessary to authenticate these models further, such work falls outside the scope of this article. Our present aim was to simply use the data to compare and contrast the effectiveness of various entropic force models.}

In summary, using quantum-deformed entropy, we could derive analytical results for the observable parameters of the effective dark energy sector, namely the density parameter and EoS parameter of the dark energy of the proposed q-deformed cosmological scenario. We look at the following in more detail in cosmological ramifications. To investigate the facts of the model, let us use the present-time best-fit value of the density parameter of cold matter.

\begin{figure}[ht]
\centering
\includegraphics[width=8cm]{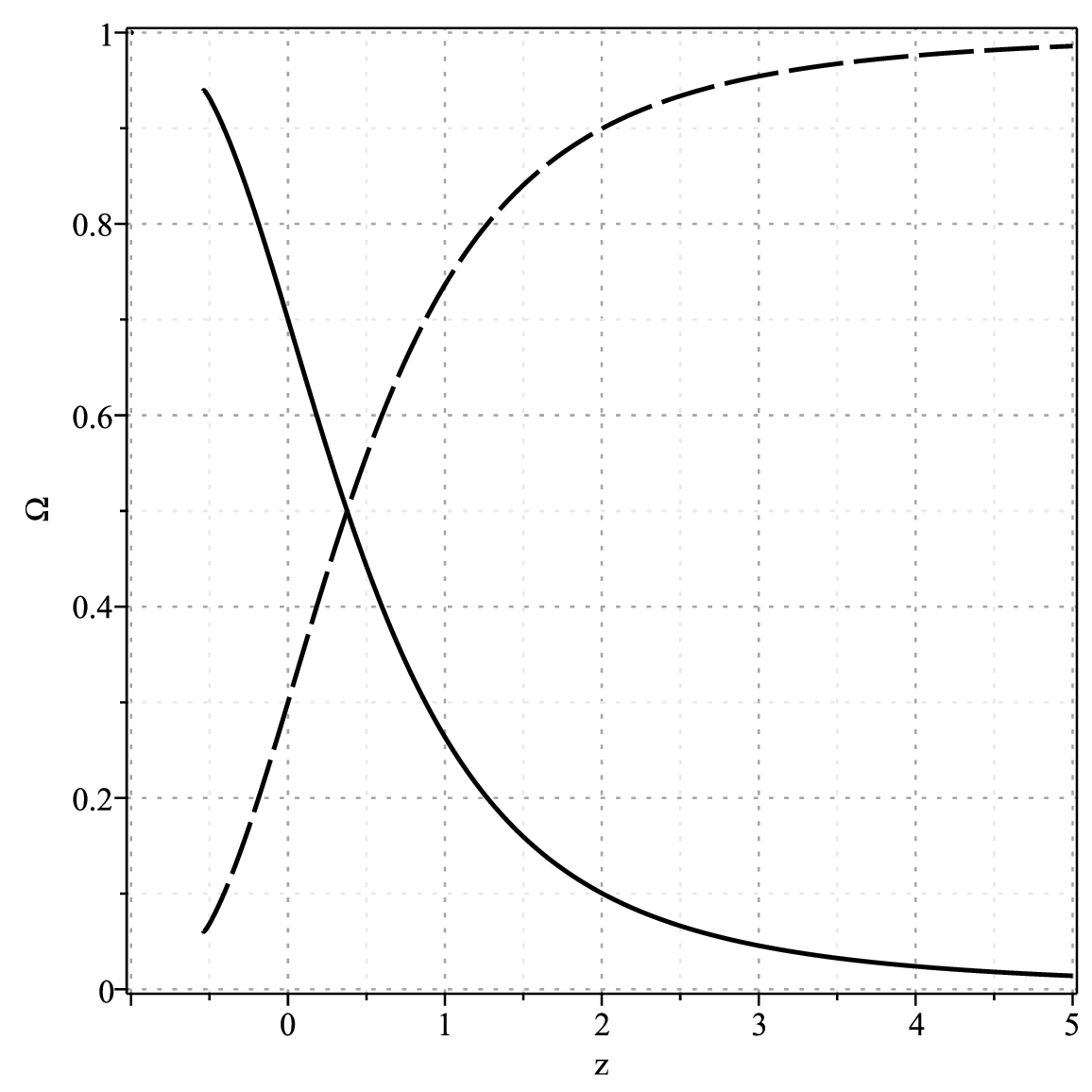}
\caption{The evolution of the effective dark energy density parameter $\Omega^\text{(DE)}$ (solid line) and the matter density parameter $\Omega^\text{(c)}$
(dashed) respectively, as a function of the redshift $z$. }\label{ppp2}
\centering
\end{figure}

{ By using the obtained values in (\ref{Fitt}), one can determine} the q-deformation parameter, $ \mathcal{N}$, and the q-deformed entropy of the apparent horizon defined by Eq. (\ref{Qentropy}). Regarding the definitions of $\mathcal N$ and $\Omega^{(q)}_0$ in Eqs. (\ref{3-14non}) and (\ref{Omegas}) we obtain
 \begin{equation}\label{NNN}
   \mathcal{N}=\frac{\Omega^{(q)}_0}{H_0^2L_\text{P}^2}=1.70\times10^{124}.
 \end{equation}
%It is worth acknowledging that Ref. \cite{Jalalzadeh:2017jdo}, with a general argument, has roughly estimated the order $\mathcal N\sim10^{120}$, a value consistent with the above outcome. In addition, Eq. (\ref{NNN}) demonstrates that the exponentiated argument in the deformation parameter $q=\exp(2\pi i/\mathcal N)$ is  extremely small. However,  as we obtained in Eq. (\ref{M2}) (see also Fig. \ref{ppp2}), its influence is crucial in the late time cosmology.
% Also, using the above value of $\mathcal N$ in Eq. (\ref{Qentropy}) gives us the value of q-deformed entropy of the apparent horizon
{It is noteworthy to acknowledge the estimations made by Refs. \cite{Jalalzadeh:2017jdo,Artymowski:2018pyg}, which through general arguments in the context of quantum cosmology, approximated the orders $\mathcal N\sim10^{120}-10^{123}$, respectively. Furthermore, Eq. (\ref{NNN}) provides evidence that the argument of trigonometric functions found in the deformation parameter, $q=\exp(2\pi i/\mathcal N)=\cos(2\pi /\mathcal N)+i\sin(2\pi/\mathcal N)$, is exceedingly minute. However, as can be observed from Eq. (\ref{M2}) (also shown in Fig. \ref{ppp2}), its impact plays a crucial role in the late time cosmology. Additionally, utilizing the aforementioned value of $\mathcal N$ in Eq. (\ref{Qentropy}) allows us to obtain the q-deformed entropy of the apparent horizon
 \begin{equation}\label{63}
     S_q(t_0)= 2.298\times10^{124}.
 \end{equation}
 Thus, the implication of the diminutiveness of $\mathcal N$ is the exceedingly high value of the entropy of the apparent horizon. Please note that in order for the universe to be of significant size, the argument of the trigonometric functions in $q$ has to be very small. Furthermore, $\mathcal N$ is representative of the entropy of the apparent horizon, as illustrated by Eq. (\ref{63}).  According to the holographic bound \cite{Susskind:1994vu}, the entropy budget of the observable universe must not surpass that of the apparent horizon. The total entropy of the observable universe is roughly $S_{obs}\simeq10^{102}-10^{103}$ \cite{2010Ap5E}, with supermassive black holes occupying a dominating role at the cores of galaxies. This also offers a hint regarding the minimum value of $\mathcal N$, which is greater than $10^{123}$, as confirmed by our discovery in Eq. (\ref{63}).}
\begin{figure}[ht]
\centering
\includegraphics[width=8cm]{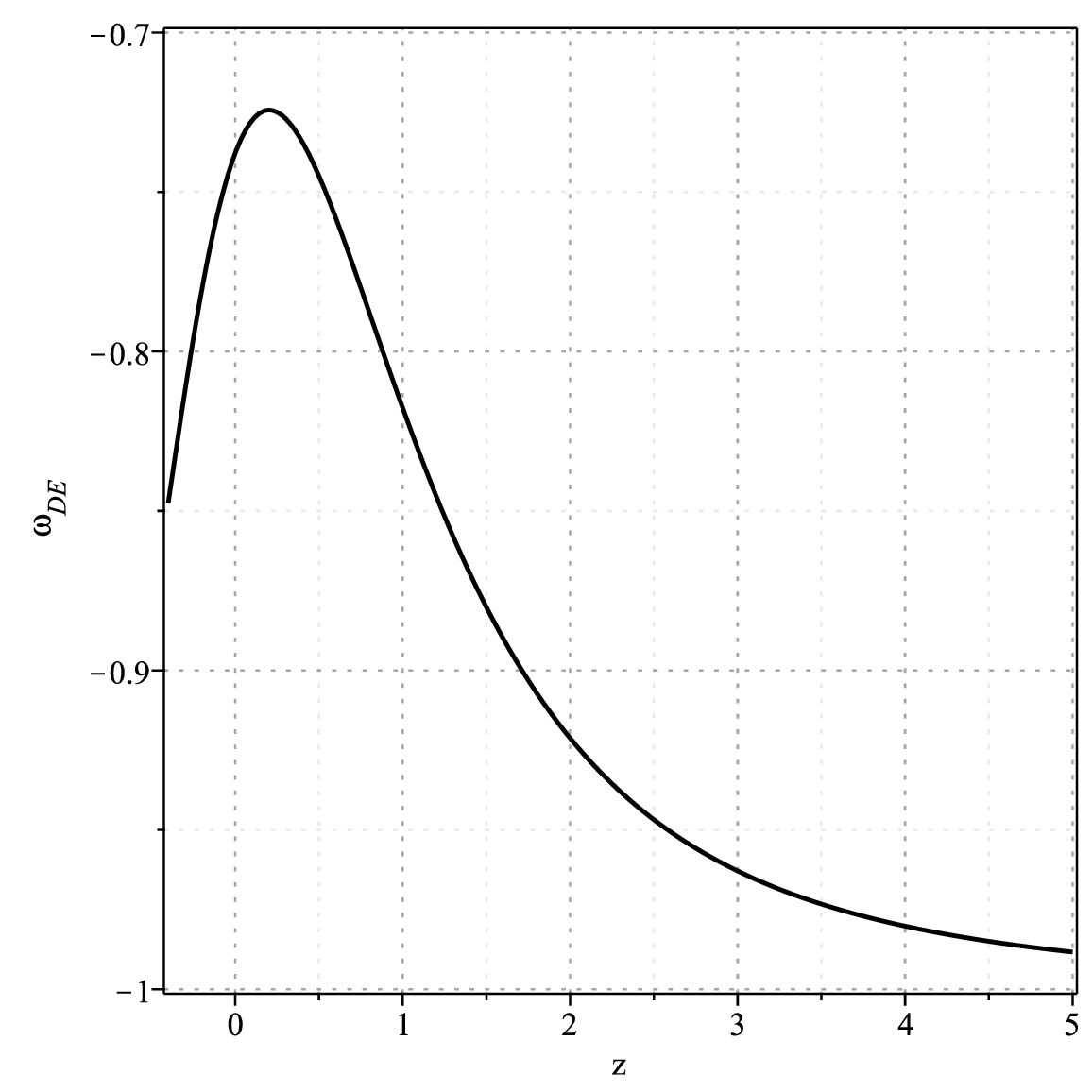}
\caption{The evolution of the effective (q-deformed) dark energy EoS
parameter, $\omega_\text{DE}$.  }\label{ppp3}
\centering
\end{figure}

Fig. \ref{ppp2} depicts the evolution of the effective dark energy, as we found in Eq. (\ref{M2}), and cold matter density parameters with redshift. It is simple to demonstrate, and as this figure indicates, at redshift $z=0.377$, two density parameters are identical, and we then have a universe dominated by dark energy. Finally, at redshift $z=-1$, the contribution of the cold matter is insignificant, resulting in a de Sitter universe. This demonstrates that the model is a plausible cosmological model since it accommodates an early matter-dominated phase in which structure could form and a recent acceleration phase corresponding to observations.

In addition, in Fig. \ref{ppp3}, we show the corresponding behavior of the effective dark energy EoS parameter as a result of (\ref{M2}). This figure shows that the EoS parameter begins at $\omega_\text{DE}=-1$ at high redshifts, achieves a maximum of $\omega_\text{DE}=-0.7243$ at redshift $z=0.2024$, and returns to $\omega_\text{DE}=-1$ at subsequent redshifts.

\begin{figure}[ht]
\centering
\includegraphics[width=8cm]{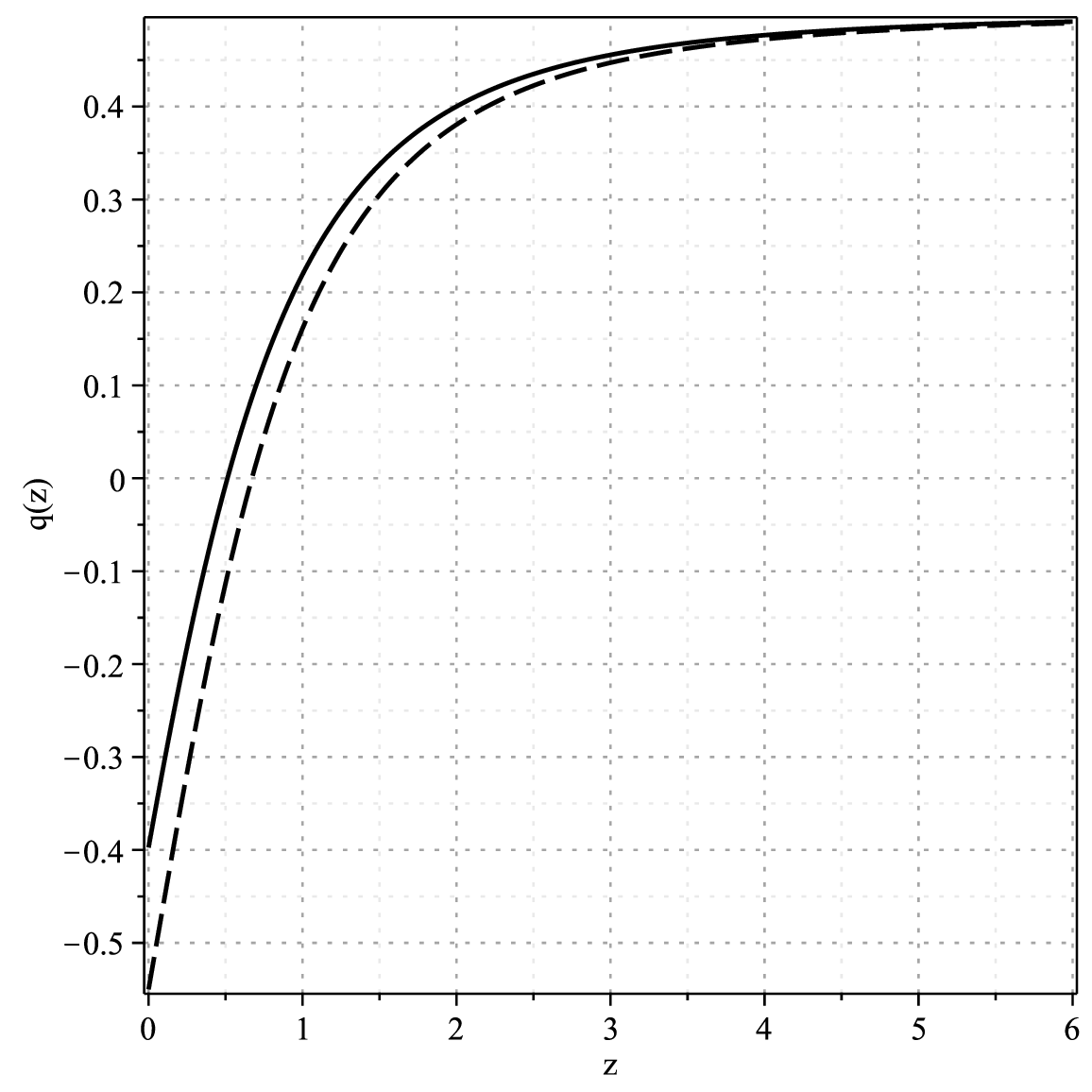}
\caption{Evolution of the deceleration parameter $ \mathfrak{q}(z)$ versus redshift. The black line exhibits the deceleration parameter of the q-deformed cosmology, with $\Omega_0^\text{(q)}=0.43$, $\Omega_0^\text{(c)}=0.3$ and  $z_\text{tran}=0.5$. The dash-line denotes the deceleration parameter of the standard flat $\Lambda$CDM model with $\Omega^{(\Lambda)}_0=0.7$ (with $z_\text{tran}=0.67$).   }\label{ppp4}
\centering
\end{figure}

Finally, we show the deceleration parameter in Fig. \ref{ppp4} defined by
\begin{equation}
    \label{q1}
    \mathfrak{q}=-1-\frac{\dot H}{H^2}=-1+\frac{3f(z)\Omega^{(q)}}{2\Omega^{(q)}_0}\left\{1+\frac{\pi\Omega^{(q)}}{4}\right\}.
\end{equation}
For the reader's convenience, we have extended evolution into the far future, i.e., $z\rightarrow-1$. This graph shows the transition from deceleration to acceleration at $z_\text{tran}=0.5$, consistent with the cosmological observations \cite{Muthukrishna:2016evq,2012JCA}.
\begin{figure}[ht]
\centering
\includegraphics[width=8cm]{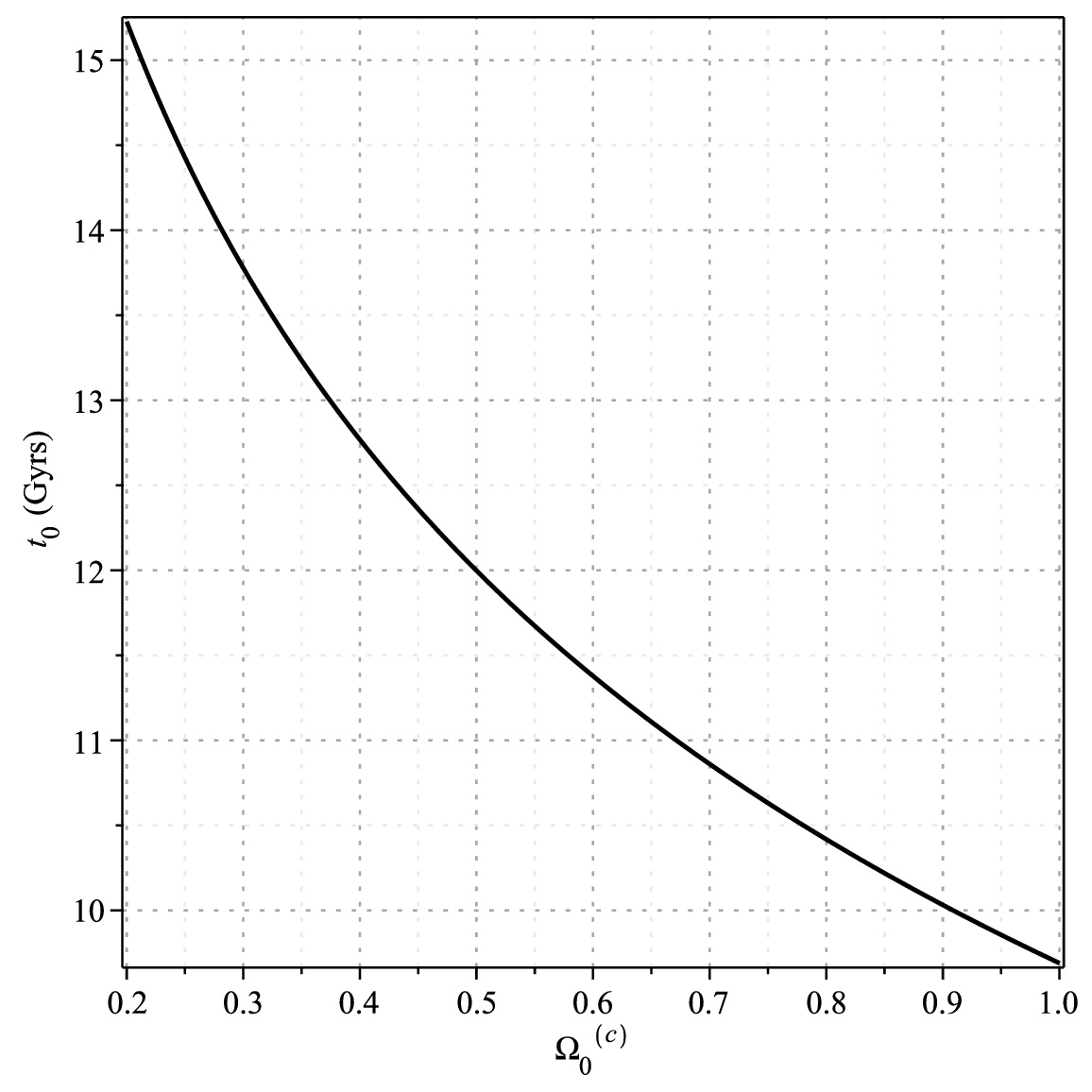}
\caption{The age of the Universe as a function of $\Omega^\text{(cm)}$. Regarding  the Planck 2018 collaboration observations, we assumed $ H_0=67.27~ \text{Km}\cdot \text{sec}^{-1}\cdot\text{Mpc}^{-1}$. }\label{ppp5}
\centering
\end{figure}
Also, Eq. (\ref{q1}) gives $ \mathfrak{q}_0=-0.3973$ for the deceleration parameter today, which is consistent with model-independent observations \cite{Muthukrishna:2016evq}.

We close this section by calculating the Universe's age according to the scenario. Figure \ref{ppp5} illustrates the Universe's age as a function of the density parameter of the cold matter. There are two different estimates for the age of the Universe. Based on early observations of the Universe, the first estimate suggests that the Universe is 13.787 billion years old according to the $\Lambda$CDM  model as of 2021 \cite{Planck:2018vyg}. The second estimate is based on observations of the current universe and suggests that the Universe is younger \cite{Riess:2018byc,Freedman:2019jwv}. Recent studies have lowered the uncertainty of the first type of measurement to 20 million years. This was achieved through various studies that produced similar results, including analyses of the microwave background radiation by the Planck satellite, the WMAP Probe, and other space missions \cite{2013ApJS20B}.

Inserting the obtained Hubble parameter in (\ref{Hubble}) to the expression of the Universe's age
\begin{equation}
    \label{age}
    t_0=\int_0^\infty\frac{dz}{(1+z)H(z)},
\end{equation}
we find, 
\begin{equation}
    \label{age1}
    t_0=\frac{0.9479}{H_0}=13.779~\text{Gyrs}.
\end{equation}
The above value coincides with the value corresponding
to the standard $\Lambda$CDM scenario, namely $13.787\pm0.020$ Gyrs \cite{Planck:2018vyg}.

\section{Comparison with other models}\label{sec5}

{The conventional Friedmann and Raychaudhuri equations involving ordinary matter fields are inadequate for characterizing the dark energy epoch of the present universe. It is in this context that we can fit the following comment. Concretely, 
Jacobson \cite{Jacobson:1995ab} successfully obtained the Einstein field equations from the entropy and horizon area proportionality by presuming the heat flow across the horizon, and Padmanabhan \cite{Padmanabhan:2003pk} derived the Friedmann and Raychaudhuri equations using the holographic equipartition law, which indicates that the difference between the degrees of freedom at the surface and in the bulk of an area of space causes cosmic space expansion. Verlinde \cite{Verlinde:2010hp} has also proposed a new concept that defines gravity as an entropic force arising in a system due to the statistical tendency to increase its 
entropy\footnote{{This theory is a significant breakthrough in the field of physics and a promising avenue for future research. All the innovative ideas above offer fresh insights into the quantum gravity puzzle and could even shed light on the thermodynamic origins of space-time. These breakthroughs in the field may significantly enhance our understanding of the universe.}}. Consequently, alternative forms of entropy, especially non-extensive entropies, distinct from the Bekenstein--Hawking variety, have been advocated to induce accelerated expansion, as now found in the literature for thermodynamic interpretations of modern cosmology.
In this regard, various non-extensive entropies have been proposed, such as Tsallis \cite{Tsallis:1987eu}, Renyi \cite{renyi1961}, Sharma--Mittal \cite{sharma1975new}, Kaniadakis \cite{Kaniadakis:2005zk},  Loop Quantum Gravity \cite{Majhi:2017zao,Liu:2021dvj}, Tsallis--Cirto \cite{2013EPJCT}, Tsallis--Zamora \cite{Zamora:2022cqz}, Barrow \cite{Barrow:2020tzx}, and fractional \cite{Jalalzadeh:2021gtq,Jalalzadeh:2022uhl} entropies.}

{Tsallis entropy is a generalized form of Gibbs/Shannon entropy applicable in cases where entropy's additive and extensive properties do not hold. Renyi entropy, on the other hand, is a measure of the entanglement of quantum systems and is commonly used in the context of BHs and cosmological horizons. Tsallis-Cirto entropy, which has been motivated by the need to make BH entropy extensive, is similar in nature. Regarding mathematical equivalence, Barrow and fractional entropies are comparable to Tsallis--Cirto entropy. However, they are primarily driven by the fractal structure of the horizon caused by quantum fluctuations. The Barrow entropy, proposed as a toy model for the potential effects of quantum gravitational spacetime foam, was not supported by any concrete evidence. Barrow argued that quantum-gravitational effects could introduce intricate, fractal features on the BH structure, similar to the illustrations of the Covid-19 virus. In contrast, the fractal entropy of black holes in fractional quantum gravity has been proven to be real, and it has been shown that the structure of the BH surface horizon is a {\it random fractal surface} distinct from Barrow's inspiration. Furthermore, the Sharma--Mittal entropy can extend the Tsallis and Renyi entropies, whereas the Kaniadakis entropy is derived from the principles of special relativity.}

{Referring to the findings presented in Ref. \cite{Nojiri:2022dkr}, it can be observed that a comprehensive overview of the majority of these entropies can be effectively encapsulated in the subsequent form
\begin{multline}
S_\mathrm{g}(\alpha_+,\alpha_-,\beta,\gamma) =\\ \frac{1}{\gamma}\left[\left(1 + \frac{\alpha_+}{\beta}~S_\text{BH}\right)^{\beta} 
 - \left(1 + \frac{\alpha_-}{\beta}~S_\text{BH}\right)^{-\beta}\right],
\label{gen-entropy}
\end{multline}
where $\alpha_+$, $\alpha_-$, $\beta$, and $\gamma$ are real and positive the parameters positive, and $S_\text{BH}$ is the Bekenstein--Hawking entropy. 
As authors of \cite{Nojiri:2022dkr} showed, by suitable choice of these parameters, the generalized entropy (\ref{gen-entropy})
reduces to the 
\begin{equation}\label{ABZ1}
    \begin{split}
           & S_g=S_\text{BH}^\beta,\hspace{2cm}\text{ (Tsallis--Barrow--Fractional)},\\
    &    (\alpha_+ \rightarrow \infty, \alpha_- = 0, \gamma = \left(\alpha_+/\beta\right)^{\beta}),\\
     &  S_g=  \frac{1}{\alpha}\ln{\left(1 + \alpha~S_\text{BH}\right)},\hspace{2cm}\text{(Rényi)}\\
    &  (\alpha_- = 0,~\beta \rightarrow0,~ \frac{\alpha_+}{\beta} \rightarrow \mathrm{finite}),\\
 &  S_g =   \frac{1}{\gamma}\left[\left(1 + \frac{\alpha_+}{\beta}~S\right)^{\beta} - 1\right],~~~\text{(Sharma--Mittal)},\\
  &    (\alpha_- = 0),\\
  & S_g= \frac{1}{K}\sinh{\left(KS\right)},\hspace{2cm}\text{(Kaniadakis)},\\
  &(\beta \rightarrow \infty, \alpha_+ = \alpha_- = \frac{\gamma}{2} = K),\\
  &  S_g=\frac{1}{(1-q)}\left[\mathrm{e}^{(1-q)S} - 1\right],~~~\text{( Loop Quantum Gravity)}\\ 
     & (\alpha_- = 0, \beta \rightarrow \infty, \gamma = \alpha_+ = (1-q)),
    \end{split}
\end{equation}
entropies.
The entropy function described in Eq. (\ref{gen-entropy})
 and the q-deformed entropy (\ref{Qentropy}) share several similar properties. Firstly, both of these functions satisfy the generalized third law of thermodynamics. Secondly, they both exhibit a monotonically increasing behavior concerning the Bekenstein--Hawking entropy. Finally, these functions converge to the Bekenstein--Hawking entropy under certain parameter limits. As previously stated in the introduction, one of the main challenges such models face is the presence of free parameters with varying origins. Upon reviewing the generalized entropy introduced in (\ref{gen-entropy}), it is apparent that the existing parameters have distinct origins. Even if the model were to provide a highly accurate representation of observational data, comprehending the physics behind these parameters would prove to be more challenging than the initial issue of the cosmological constant problem in the standard model of cosmology. }

{ If we use the generalized entropy (\ref{gen-entropy}) in process of section \ref{FR}, the resulting  Friedmann equation will be \cite{Nojiri:2022dkr}
\begin{multline}\label{Var}
\frac{G\beta H^4}{\pi\gamma}\Bigg\{ \frac{1}{\left(2+\beta\right)}\left(\frac{GH^2\beta}{\pi\alpha_-}\right)^{\beta}~
_2F_{1}\Big(1+\beta, 2+\beta, 3+\beta,\\ -\frac{GH^2\beta}{\pi\alpha_-}\Big) + \frac{1}{\left(2-\beta\right)}\left(\frac{GH^2\beta}{\pi\alpha_+}
\right)^{-\beta}~_2F_{1}\Big(1-\beta, 2-\beta,\\ 3-\beta, -\frac{GH^2\beta}{\pi\alpha_+}\Big) \Bigg\} = \frac{8\pi G\rho}{3} + \frac{\Lambda}{3},
\end{multline}
where $_2F_1(a,b,c,d)$ denotes the Hypergeometric function. Note that the ``cosmological constant'', $\Lambda$, appears in the above generalized Friedmann equation as an integration constant. This is where our model truly distinguishes itself, as it separates from the generalized entropy and its various subset entropies as defined in equation (\ref{gen-entropy}). Let us go through this in further depth. The CC seen in the previous equation is a constant of integration that can be chosen arbitrarily. In contrast, the constant of integration in the Friedmann equation (\ref{Freed}) is fixed to a specific value. Understanding this fixed value is simple. As the universe approaches very late times and the scale factor tends towards infinity, the energy density of matter fields in the right-hand side of the equation vanishes. Additionally, the first term on the right-hand side of (\ref{Freed}), which is $\frac{\cos(\frac{\gamma}{G}R_\text{AH}^2)}{R_\text{AH}^2}$, identically vanishes as well. This boundary condition fixes the value of the constant of integration to $\frac{\pi\Lambda_q}{6}\Si\left(\frac{\pi}{2}\right)$.}

{The presence of the CC in the generalized Friedmann equation leads to the CC problem. The challenge is explaining why the measured CC is not precisely zero but has a very small nonzero value. In contrast, the q-deformed Friedmann equation (\ref{Freed}) does not suffer from this problem.}

{We can use Barrow's entropy model as an illustrative example to provide a more precise reference point. In the context of cosmology, various perspectives have been explored regarding the impact of Barrow entropy on the universe's evolution. One such view involves modifying the area law, which results in a novel holographic dark energy model based on Barrow entropy \cite{Saridakis:2020zol,Anagnostopoulos:2020ctz}. Another cosmological scenario that incorporates Barrow entropy was presented in Ref. \cite{Saridakis:2020lrg}, where it was demonstrated that new additional terms emerge in the Friedmann and Raychaudhuri equations, forming an effective dark energy sector. While Ref. \cite{Saridakis:2020lrg} argued that the modified cosmological equations rooted in Barrow entropy (\ref{ABZ1}) could account for the thermal history of the universe, spanning from early deceleration to later acceleration during the dark-energy epoch that follows the cold matter dominated epoch, regardless of the presence of CC, $\Lambda$, it seems that this conclusion is only accurate if there is a CC present \cite{Saridakis:2020lrg,Sheykhi:2022jqq}. 
To illustrate this important point, let us rewrite the Friedmann and Raychaudhuri equations with Barrow's entropy \cite{Saridakis:2020lrg,Sheykhi:2022jqq}
\begin{equation}
    \begin{split}\label{Shey}
  &    H^2=\frac{8\pi G}{3}(\rho+\rho_\text{DE}),\\
  &     \dot H=-4\pi G(\rho+p+\rho_\text{DE}+p_\text{DE}),
    \end{split}
\end{equation}
where $\rho$ and $p$ are the energy density and pressure of the ordinary matter, and
\begin{equation}
    \begin{split}
        \rho_\text{DE}&=\frac{3}{8\pi G}\Big\{\frac{\Lambda}{3}+H^2\Big[1-\frac{\Delta+2}{2-\Delta}\left(\frac{\pi}{G}\right)^\frac{\Delta}{2}H^{-\Delta}\Big]\Big\},\\
        p_\text{DE}&=-\frac{1}{8\pi G}\Big\{\Lambda+2\dot H\Big[1-(1+\frac{\Delta}{2})\left(\frac{\pi}{G}\right)^\frac{\Delta}{2}H^{-\Delta}\Big]\\
       & +3H^2\Big[1-\frac{\Delta+2}{2-\Delta}\left(\frac{\pi}{G}\right)^\frac{\Delta}{2}H^{-\Delta} \Big]\Big\},
    \end{split}
\end{equation}
where  $\Delta=2(\beta-1)$, $(0\leq\Delta\leq1)$ is the Barrow's deformation parameter. For $\rho=p=0$ and $\Lambda=0$, the solution of the above field equations is $H=0$. On the other hand, for $\rho=p=0$ and $\Lambda\neq0$ we obtain
\begin{equation}
    H=\left(\frac{\Lambda}{3}\right)^\frac{1}{2-\Delta}.
\end{equation}
These findings validate that the accelerated phase can solely occur when the CC is present. However, we now have two unknown parameters in the theory, $\Lambda$ and $\Delta$. As previously discussed, the presence of a constant of integration, $\Lambda$, reintroduces the cosmological constant problem and warrants an explanation. Furthermore, it is imperative to provide a sound justification for the particular numerical value of $\Delta$ that is deemed valid based on cosmological observations.}

{On the other hand, the q-deformed Friedmann and Raychaudhuri equations (\ref{Freed1}) and (\ref{Raych}) for $\rho=p=0$ reduce to
\begin{equation}
    \label{Freed100}
   \cos(\frac{\pi\Lambda_q}{6H^2})H^2 =\frac{\pi\Lambda_q}{6}\left\{ \Si\left(\frac{\pi}{2}\right)-\Si\left(\frac{\pi\Lambda_q}{6H^2}\right)\right\},
\end{equation}
and
\begin{equation}
    \label{Raych00}
    \cos(\frac{\pi\Lambda_q}{6H^2})\frac{\ddot a}{a}=\frac{\pi\Lambda_q}{6}\left\{ \Si\left(\frac{\pi}{2}\right)-\Si\left(\frac{\pi\Lambda_q}{6H^2}\right)\right\}.
\end{equation}
The solution of these equations is a de Sitter spacetime, $H^2=\Lambda_q/3$, in which $\Lambda_q$ plays the role of the CC. As previously described, our initial cosmological model did not include a conventional CC. However, our subsequent implementation of quantum deformation resulted in the emergence of an effective CC, which is strongly linked to the natural number $\mathcal{N}$ defined by equation (\ref{3-14non}). This constant is directly derived from the assumption of a finite number of states in Hilbert spaces, as stated in \cite{Jalalzadeh:2017jdo}.}

\section{Conclusions}\label{Con}

In this paper, we developed a novel cosmological scenario 
assuming that thermodynamics is intertwined
with gravity. It is known, for instance, that one may start with the first law of thermodynamics, apply it to the Universe horizon, and end up with the Friedmann and Raychaudhuri equations.
This method utilizes the Bekenstein--Hawking entropy in the case of GR or the modified entropy expression in the case of modified gravity. 

We examined the compatibility of the apparent horizon's entropy with thermodynamic laws on the assumption that it is the q-deformed entropy of the black hole-white hole pair. 
To achieve this, we first presupposed that the apparent horizon is subject to the unified first law of thermodynamics and that its entropy has the form (\ref{hoo1}). Then we demonstrated how the unified first law of thermodynamics on the apparent horizon might be expressed as modified Friedmann equations of an FLRW universe with arbitrary spatial curvature.

Our research shows that the q-deformed Friedmann and Raychaudhuri equations lead to the existence of an effective dark energy component, which can explain the Universe's late-time acceleration. According to our model, the density parameters of cold matter and the effective dark energy are equal at a redshift of $z=0.377$, after which the effective dark energy dominates the universe. Additionally, our model predicts a de Sitter universe in the long run. The model predicts that the effective dark energy equation of state (EoS) parameter changes over time. Initially, it starts at $\omega_\text{DE}=-1$ during high redshifts. Then, it reaches a peak at $\omega_\text{DE}=-0.7243$ at redshift $z=0.2024$. After that, it returns to its initial value of $\omega_\text{DE}=-1$ during subsequent redshifts. Furthermore, we find a transition from deceleration to acceleration at $z_\text{tran}=0.5$, consistent with the cosmological observations. The obtained age of the Universe is 13.779 Gyrs, which coincides with the value corresponding to the standard $\Lambda$CDM scenario, namely $13.787\pm0.020$ Gyrs. 
These findings suggest that our model is a viable cosmological model, as it accommodates an early phase of matter domination that allowed the formation of structures and a recent acceleration phase that aligns with observations.
{Concretely,  we support this claim upon using suitable calculations and corresponding plots, in particular of  $H/H_0$ (radiation plus cold matter) versus the scale factor, either 
in flat $\Lambda$CDM or our q-deformed cosmology, and the evolution of luminosity distance, $\mu$, versus redshift $z$, the evolution of $H(z)$ versus redshift $z$ (including error bars) and the evolution of the deceleration parameter $q(z)$ versus redshift (see figures in section \ref{Cosmos}).
Additionally, 
we have conducted a thorough comparative analysis of our proposed model with others involving non-extensive entropies.}

In conclusion, the q-deformed horizon entropy cosmology scenario demonstrates intriguing phenomenology and is consistent with cosmological observations. As a result, it may be an intriguing option for the description of Nature.

{Last but not least, let us add that it is imperative to acknowledge that the current models possess both advantageous and disadvantageous features and are not without flaws. 
For example, in entropic cosmology, the form of the driving entropic force terms is determined by the definition of entropy. In the original entropic force models, the Bekenstein entropy and the Hawking temperature have been used to get the entropic force terms. However, such models fail to account for both the Universe’s acceleration and deceleration, and it has been shown in \cite{Basilakos:2012ra,Basilakos:2014tha} that they neither account for cosmic fluctuations nor are compatible with the formation of structures. As we have thoroughly discussed in section \ref{sec5}, using Barrow entropy-based models, with the addition of the cosmological constant as an integration constant, can effectively elucidate the shift from a decelerated universe to an accelerated universe. Nevertheless, including the cosmological constant in the model leads to a resurgence of the cosmological constant problem and the incorporation of obscure parameters into the cosmological model, further complicating the matter. Therefore, exploring alternative models that can account for these phenomena is necessary and provides a more accurate description of the Universe's behavior. Ultimately, this will lead to a better understanding of our Universe and its complex dynamics.}

%\section*{CRediT authorship contribution statement} S. Jalalzadeh: Investigation, Writing--original draft, Formal analysis, Visualization, Supervision. H. Moradpour: Writing --review, Formal analysis \& editing. P.V. Moniz: Writing --review, Formal analysis \& editing.

%\section*{Declaration of competing interest}
%The authors declare that they have no known competing financial interests or personal relationships that could have appeared to influence the work reported in this paper.

%\section*{Data availability}
%No data was used for the research described in the article.

\section*{Acknowledgements}
S.J. acknowledges financial support from the National Council for Scientific and Technological Development--CNPq, Grant no. 308131/2022-3.  {P.V.M. acknowledges the FCT grants 
UID-B-MAT/00212/2020 at CMA-UBI as well as the COST Action CA18108 (Quantum gravity phenomenology in the multi-messenger approach).}

\bibliographystyle{elsarticle-num}
\bibliography{copy}

\end{document}